\documentclass[twocolumn]{aastex62}
\usepackage{multirow}

\newcommand{\teff}{$T_{\rm eff}$}
\newcommand{\logg}{$\log g$}
\received{Jan. 1, 2018}
\revised{Jan. 7, 2018}
\accepted{\today}
\submitjournal{ApJ}

\shorttitle{effective temperatures for K8 -- M5 stars}
\shortauthors{L\'opez-Valdivia et al.}

\usepackage{hyperref}
\begin{document}

\title{Effective Temperatures of Low-Mass Stars from High-Resolution H-band Spectroscopy}

\correspondingauthor{Ricardo L\'opez-Valdivia}
\email{rlopezv@utexas.edu}

\author[0000-0002-7795-0018]{Ricardo L\'opez-Valdivia}
\affil{The University of Texas at Austin,
Department of Astronomy, 
2515 Speedway, Stop C1400, 
Austin, TX 78712-1205}

\author[0000-0001-7875-6391]{Gregory N. Mace}
\affil{The University of Texas at Austin,
Department of Astronomy, 
2515 Speedway, Stop C1400,
Austin, TX 78712-1205}

\author[0000-0002-3621-1155]{Kimberly R. Sokal}
\affil{The University of Texas at Austin,
Department of Astronomy,
2515 Speedway, Stop C1400,
Austin, TX 78712-1205}

\author[0000-0001-9580-1043]{Maryam Hussaini}
\affil{The University of Texas at Austin,
Department of Astronomy,
2515 Speedway, Stop C1400,
Austin, TX 78712-1205}

\author{Benjamin T. Kidder}
\affil{The University of Texas at Austin,
Department of Astronomy,
2515 Speedway, Stop C1400,
Austin, TX 78712-1205}

\author[0000-0003-3654-1602]{Andrew W. Mann}
\affil{Department of Physics and Astronomy, University of North Carolina at Chapel Hill, Chapel Hill, NC 27599, USA}

\author[0000-0002-8443-0723]{Natalie M. Gosnell}
\affiliation{Department of Physics,
Colorado College,
14 E. Cache La Poudre St,
Colorado Springs, CO 80903}

\author[0000-0002-0418-5335]{Heeyoung Oh}
\affil{The University of Texas at Austin,
Department of Astronomy,
2515 Speedway, Stop C1400,
Austin, TX 78712-1205}
\affil{Korea Astronomy and Space Science Institute, 776 Daedeok-daero, \\
Yuseong-gu, Daejeon 34055, Korea}

\author{Aurora Y. Kesseli}
\affil{Boston University, 725 Commonwealth Ave.,
Boston, MA 0215}

\author{Philip S. Muirhead}
\affil{Boston University, 725 Commonwealth Ave.,
Boston, MA 0215}

\author{Christopher M. Johns-Krull}
\affil{Physics \& Astronomy Dept., Rice University,
6100 Main St., Houston, TX 77005}

\author[0000-0003-3577-3540]{Daniel T. Jaffe}
\affil{The University of Texas at Austin,
Department of Astronomy,
2515 Speedway, Stop C1400,
Austin, TX  78712-1205}

\begin{abstract} 
High-resolution, near-infrared spectra will be the primary tool for finding and characterizing Earth-like planets around low-mass stars.
Yet, the properties of exoplanets can not be precisely determined without accurate and precise measurements of the host star. 
Spectra obtained with the Immersion GRating INfrared Spectrometer (IGRINS) simultaneously provide diagnostics for most stellar parameters, but the first step in any analysis is the determination of the effective temperature. 
Here we report the calibration of high-resolution H-band spectra to accurately determine effective temperature for stars between 4000-3000~K ($\sim$K8--M5) using absorption line depths of Fe~{\sc i}, OH, and Al~{\sc i}.
The field star sample used here contains 254 K and M stars with temperatures derived using BT-Settl synthetic spectra.
We use 106 stars with precise temperatures in the literature to calibrate our method with typical errors of about 140~K, and systematic uncertainties less than $\sim$120~K. 
For the broadest applicability, we present \teff--line-depth-ratio relationships, which we test on 12 members of the TW Hydrae Association and at spectral resolving powers between $\sim$10,000--120,000. These ratios offer a simple but accurate measure of effective temperature in cool stars that is distance and reddening independent.
\end{abstract}


\section{Introduction} \label{sec:intro}
Low-mass stars ($0.1M_\odot < M_{*} < 0.6M_\odot$) represent more than 70\% of the stars in the Galaxy (e.g. \citealp{reid97,bochansky10}) and approximately 40\% of the stellar mass content (e.g. 
\citealp{mera96,chabrier05}). The main-sequence lifetimes of M dwarfs, which exceed a Hubble time, makes them valuable for deciphering Galactic formation, structure, chemical evolution and dynamics. 
Lately, M dwarfs have become the preferred targets of exoplanet searches since, for the same size exoplanet, the transit depth and the reflex motion produced is greater than around solar type stars (e.g. \citealp{bonfils12,gillon16,gillon17}). 
Therefore, a precise determination of the stellar properties of low-mass dwarfs is fundamental to understanding astronomical questions in both the Galactic and planetary contexts.

Historically, effective temperature (\teff) has been determined from  photometric data (eg. \citealp{alonso96,masana06,casagrande10,hawkins16}), 
excitation equilibrium  (eg. \citealp{santos00,sousa11,santos13}), line-depth ratios (eg. \citealp{gray91,biazzo07,fukue15,taniguchi18}), and spectral fitting (eg. \citealp{prugniel11,sharma16, garcia-perez16}). 
Each of these methods have distinct applications and potential drawbacks, with the resulting temperature scales differing between them by a few-hundred Kelvin. 

For example, \cite{veeder74} and \cite{bessel91} obtained a temperature scale for M stars by fitting a blackbody to optical and near-infrared fluxes. 
The \cite{veeder74} temperature scale for stars later than M5 resulted in a much cooler sequence ($\sim$180~K) than that found by \cite{bessel91}.
\cite{casagrande08} obtained a temperature scale for M dwarfs by modifying the infrared flux method (IFM) used for FGK dwarfs \citep{casagrande06}.
The IFM relies on the assumption that the M star flux beyond $\sim$2.0~$\mu$m is approximately a blackbody. 
However, M stars have more flux than the blackbody prediction at those wavelengths \citep{rajpurohit13}, and as a consequence the IFM temperatures may be  underestimates.

When using spectra to determine \teff there is the added benefit of independent indicators for other physical properties like surface gravity and metallicity.
Nevertheless, the determination of \teff\ in low-mass stars from high-resolution infrared (IR) spectra is complicated by incomplete spectral line lists, incorrect absorption line strengths, and the presence of diatomic (e.g. TiO, FeH, OH, CO) and triatomic (e.g. H$_2$O) absorption bands.  
Despite these challenges, \cite{rajpurohit13} determined \teff\ through a $\chi^2$ minimization between low- and moderate-resolution ($\Delta\lambda$ = 10\AA\ and $\sim$4\AA) optical ($\sim$5,200--10,000~\AA) spectra and BT-Settl \citep{btsettl} synthetic spectra. 
Those optical spectra include atomic (Ca~{\sc i}, Na~{\sc i}, K~{\sc i}), diatomic (MgH, TiO, VO, CaH) and even triatomic (CaOH) absorption features. \cite{veyette17} also determined \teff, [Fe/H] and [Ti/Fe] for 29 M dwarfs, but  using Y-band high-resolution (R$\sim$25,000) spectra and equivalent widths of several lines of Fe~{\sc i}, Ti~{\sc i}, and a FeH temperature-sensitive index. 

More recently, \cite{rajpurohit18} used a $\chi^2$ minimization method and high-resolution (R=22,000) H-band spectra along with BT-Settl models to obtain \teff, surface gravity (\logg) and metallicity ([Fe/H]) for 45 M
dwarfs. Additionally, \cite{rajpurohit18b} used optical and near infrared ($\sim$7,500--17,000~\AA) high-resolution (R=90,000) spectra to determine the stellar parameters of 292 M stars, through a $\chi^2$ minimization against BT-Settl models for certain wavelength regions, which includes Ti~{\sc i}, Fe~{\sc i}, Ca~{\sc ii}, Na~{\sc i} and OH lines. 
Rajpurohit et al. found a systematic offset between their determinations and those of \cite{passegger18}, using the same spectra, of about 200--300~K. \cite{passegger18} used $\gamma$--TiO band, a few atomic lines (Fe~{\sc i}, Ti~{\sc i}, Ca~{\sc i}, Mg~{\sc i}) and PHOENIX-ACES \citep{husser13} models to determine \teff, \logg\ and [Fe/H]. 
Since both \cite{rajpurohit18b} and \cite{passegger18} used the same spectra, the discrepancy shows that \teff\ determinations are still model-dependent. 
Such model-dependency can be corrected for by calibrating against empirical temperatures to obtain a calibrated temperature sequence. 

Stars with physical parameters constrained by interferometric observations help to mitigate model-dependency by calibrating relationships between \teff\ and stellar radius. For example \cite{mann132} derived relations between temperature sensitive indexes in the visible, J, H and K bands and \teff, \cite{newton15} used equivalent widths of some H-band temperature sensitive features (Mg, K, Si, CO and Al) to derive relations between \teff, radius and luminosity. 
\cite{mann15} used spectrophotometric calibrations to derive \teff, stellar radius, among other stellar parameters. 
The works of \cite{mann13,mann15} and \cite{newton15} used 20+ stars with interferometric measurements to calibrate their model-independent relationships with $\sim$150~K precision.

In this paper, we present the determination of \teff\ from high-resolution (R$\sim$45,000) H-band spectra, obtained with the Immersion GRating INfrared Spectrometer \citep[IGRINS;][]{yuk10,park14} for 254 K and M dwarf stars. 
Our temperature scale is calibrated with the (r - J) color-Temperature relation from \cite{mann15}. 
We also investigate the influence of \logg, projected rotational velocity ($v \sin i$), and [Fe/H] on our final results. 
Finally, we present \teff--line-depth ratios relationships that could theoretically extend our method to any H-band spectrum with resolution $>$10,000.

\section{Observations and Data Reduction} \label{sect:obs}
This analysis makes use of spectra of K and M stars observed with IGRINS since commissioning in 2014 on the 2.7~m Harlan J. Smith Telescope (HJST) at McDonald Observatory, the 4.3~m Discovery 
Channel Telescope (DCT) at Lowell Observatory, and the 8.1~m Gemini South Telescope. 
IGRINS has no moving parts and the spectral format is fixed, with R$\sim$45,000 over the entire H and K bands (14,500 to 24,500~\AA) \citep{mace16,mace18}.
Changes to the input optics ensure that the spectrum is unchanged at each facility and our analysis is homogeneous. 

We began with all $\sim$4,900 IGRINS observations between 2014 July and 2018 July. 
Based on object name and coordinates, spectral types (SpT) and literature photometry for the entire sample were obtained from the SIMBAD database \citep{wenger00} in January 2019. 
The large list of references and methodologies used to assign the spectral types listed in SIMBAD result in spectral type uncertainties of $\pm$1-2 subtypes. Spectral types were used in our analysis to provide an initial estimate of \teff\ and guide the search for atomic/molecular lines sensitive to changes in \teff\ and then to provide a \teff--SpT relation.
Giant and young stars were removed from further consideration through photometric selection using M$_K$ magnitudes derived from 2MASS photometry and Gaia DR2 parallaxes. We find that giants have M$_K$~$<$0, and YSOs are more than 1 magnitude brighter than the field M dwarf trend identified by \citet{mann15}. Such selection criteria did not rid our sample of binary stars, especially in cases where the component masses and fluxes differ significantly, and there is a possibility that our sample includes single- and double-lined spectroscopic binaries. 
The final sample we consider contains 254 stars (41 K, 198 M and 15 unknown spectral types) with 2MASS H-band magnitudes from 3 to 12.
Many of the 254 stars in this sample are well known field stars included in the analyses of \cite{mann15, rojas-ayala12, newton15, mann18b} and presented previously in the IGRINS Spectral Library \citep{Park2018}\footnote{\url{http://starformation.khu.ac.kr/IGRINS_spectral_library}}. 
 
We observed each star in our sample by nodding between two positions on the slit to facilitate the removal of sky background and telluric emission lines in data reduction. Single frame exposure times range from 30 to 900~s with the goal of achieving SNR$\gtrsim$100 per resolution element for each observation, however, 85 objects in our sample have SNR less than 100 due to conditions at the time of the observations and/or the faintness of the star. The average SNR for the sample is $\sim$160. 
A0V standard stars were observed at a similar airmass before or after each science object and used for telluric correction.

All the spectroscopic data were reduced using the IGRINS pipeline \citep*{lee17}\footnote{\url{https://github.com/igrins/plp/tree/v2.1-alpha.3}}, which performs flat-field correction, wavelength calibration using night sky OH emission and telluric absorption lines, A-B frame subtraction to remove skyline emission, and the extraction of the one-dimensional spectrum following the optimal methods of \cite{horne86}. 
Telluric absorption lines were corrected by dividing the science spectrum by the A0V spectrum, which had been multiplied by the Vega model of \cite{kurucz79}. 
A representative sample of the IGRINS spectra in our sample is shown in Figure~\ref{fig:regions}.

\section{Spectral analysis}\label{sect:spec}

Stellar spectra are primarily shaped by \teff, \logg\ and [Fe/H]. When deriving these parameters using high-resolution spectra, stellar activity, $v \sin i$ and magnetic field strength ($B$) should also be considered. 

To identify temperature sensitive spectral regions in the IGRINS spectra we first sorted the spectra by the literature spectral types.
We estimated the radial velocity of each star by finding the wavelengths offset of the Na~{\sc i} doublet at $\sim$22056 and 22084~\AA, and then we shifted all spectra to the same rest-frame wavelength. 
This process assumed that all the stars in the sample have roughly the same \logg\ and [Fe/H].
Through visual inspection we identified some new regions with \teff\ sensitivity and spectral regions that have been previously used by similar studies \citep[eg.][]{prato02, garcia-perez16, rajpurohit18}. 
We ultimately selected strong absorption lines that were close enough to each other to reside in the same IGRINS spectral order and that displayed opposite line strength variation versus SpT (\teff) (i.e. one line grew weaker and the other grew stronger when looking at progressively later spectral types).
Finally, we repeated the visual inspection using synthetic spectra and selected lines with low sensitivity to changes in \logg\ or [Fe/H].
From our visual inspection process we identified two spectral regions, bounding OH (15600 -- 15650~\AA) and Aluminum (16700 -- 16780~\AA) absorption features, that reliably trace \teff.
  
The determination of $v \sin i$ for the IGRINS spectra relied on the code developed by \citet{kesseli18}.
In spectral type bins of K0-K3, K4-K6, K7-K9, M0-M1, M2-M3, M4-M5, M6-M9 we identified template objects by their narrow lines and high signal-to-noise ratios. We selected HD~88925, HD~122120, GJ~169, GJ~15A, GJ~725A, GJ~15B, GJ~412B as our template stars for each of the spectral type bins listed above, respectively. 
We were able to determine $v \sin i$'s spanning between 7 and 53~km s$^{-1}$ for 156 stars of our sample, with $\sim$56\% between 7 and 12~km s$^{-1}$. 
The remaining stars have $v \sin i$ below the IGRINS spectral resolution and were assigned $v \sin i = 7$~km s$^{-1}$.

\subsection{Synthetic spectra}

Once the OH and Al regions were identified as the best \teff\ indicators in the IGRINS spectra of K and M stars, we looked for a theoretical counterpart (synthetic spectra) suitable for assigning temperatures. 
The BT-Settl models \citep{btsettl,baraffe15} have previously been validated in the range 2500~$\leq$~\teff~$\leq$~4000~K at low \citep[$\Delta \lambda$ = 10\AA; e.g.][]{rajpurohit13} and high \citep[$R =$ 22,000 and 90,000; e.g.][]{rajpurohit18, rajpurohit18b} spectral resolution, and is the preferred set of synthetic spectra for our study.
We employed the CIFIST\footnote{\url{https://phoenix.ens-lyon.fr/Grids/BT-Settl/CIFIST2011_2015/}} version, which cover  
the parameter space \teff~=~300~--~7000~K, \logg~=~2.5~--~5.5, [Fe/H]~=~$-$2.5~--~0.0 at high-resolution (R$\sim$330,000 at 16500~\AA). These set of spectra were computed with the {\sc phoenix} code \citep{phoenix}, the \cite{caffau11} solar abundances and an updated atomic and molecular line opacities \citep[see][and references therin]{baraffe15}, which dominate the optical and near-infrared spectra of cool stars. 

The synthetic spectra (or model) grid used in this work spans \teff\ between 2000 and 4700~K in steps of 100~K, \logg\ = 4.0, 4.5, and 5.0, solar metallicity and no alpha-element enrichment. The resolution of the synthetic spectra were degraded to the IGRINS spectral resolution ($\sim$45,000).
For all temperature determinations, we selected models with \logg\ of 4.5 since it is suitable for both K (e.g. \logg~$\sim$~4.4$\pm$0.1; \citealp{sousa08,tsantaky13}) and M (e.g. \logg~$\sim$~4.8$\pm$0.2; \citealp{segransan03,berger06}) field stars. The remaining models
with \logg\ of 4.0 and 5.0 were employed just to assess the impact of \logg\ on our analysis. 
The grid of synthetic spectra was broadened to the rotation velocities encompassing the IGRINS sample $v \sin i$'s (7 to 55 km s$^{-1}$) using the function {\tt rotBroad}, available in the PyAstronomy library\footnote{\url{https://github.com/sczesla/PyAstronomy}}. 
The rotational broadening kernel requires a linear limb-darkening coefficient, which we estimated by comparing the model \teff\ and \logg\ to \cite{claret12}\footnote{We 
used the filter H (2MASS) linear limb-darkening coefficients. For those \teff\ and \logg\ values that were not reported in \citet{claret12}, we have used the nearest (in terms of \teff\ and \logg) coefficient available, in those cases where there were more than one possible coefficient we assigned an average.} catalog. Finally, vacuum wavelengths provided with BT-Settl spectra were converted to their corresponding air wavelengths following the IAU standard formulation \citep{morton91}. 

In summary, the grid of synthetic spectra used for measuring line-depths in the OH and Al regions had \teff~=~2000~--~4700~K, \logg~=~4.5, [Fe/H]~=~0.0, $v \sin i$~=~7~--~55~km~s$^{-1}$, spectral resolution of 45,000 and no $\alpha$-element enrichment.
Figure~\ref{fig:conti} shows how the line-depth behavior in the IGRINS spectra is well reproduced by the BT-Settl models, including the flux peak in the Al region (right panel of Figure~\ref{fig:ld}). 

%

\subsection{OH region (15600 -- 15650~\AA)}
\begin{figure*}
\includegraphics[width=\textwidth]{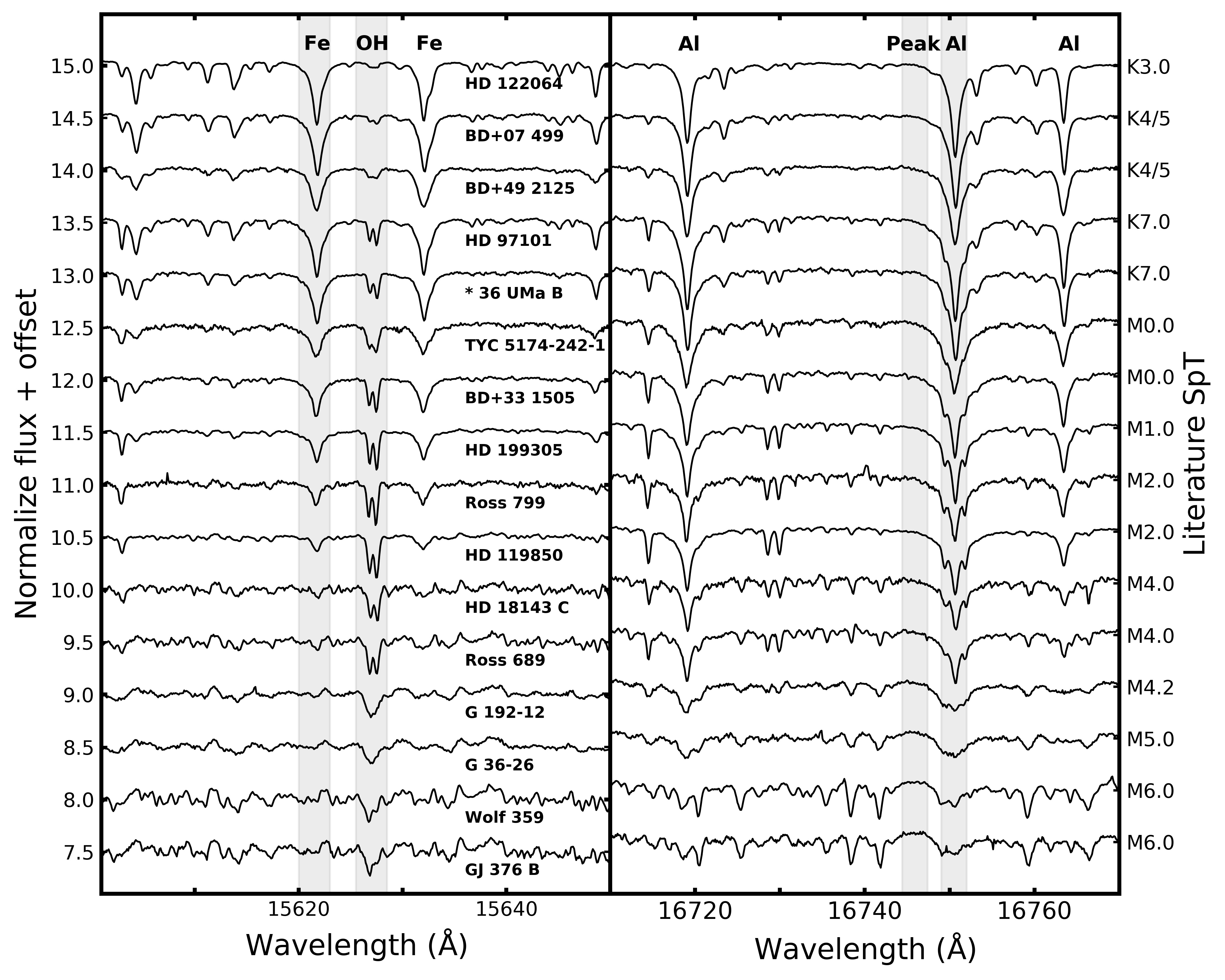}
\caption{A representative sample of IGRINS K and M star spectra around the OH region and the central 60~\AA\ of the Al region, as a function of  spectral type. The spectral lines used in this work are highlighted in gray, while other prominent lines are also labeled. The dependency of the selected lines with SpT (\teff) is clearly present. The Al region is effective for stars later than $\sim$M4 while the OH region is effective for stars earlier than $\sim$M4.\label{fig:regions}}
\end{figure*}

The OH region spans 15600 to 15650~\AA\ and includes two Fe~{\sc i} lines ($\lambda$ $\sim$15621.6 and 15631.9~\AA) and an OH ($\lambda_{\rm OH} \sim 15627.0$~\AA) doublet. 
These lines change as a function of spectral type \citep{prato02} as can be seen in the left panel of Figure~\ref{fig:regions}. 
The Fe~{\sc i} line that we used here ($\lambda_{\rm Fe}= 15621.6$~\AA) becomes weaker at lower temperatures and is un-blended in the temperature region we are interested in. 
The OH feature, which is formed by two OH lines at approximately 15626.7~\AA\ and 15627.5~\AA, increases in depth at lower temperatures,  up to  $\sim$M4--M5  stars, where numerous H$_2$O features start to dominate the spectral region.

\subsection{Al region (16700 -- 16780~\AA)}
The Al region covers 16700 to 16780~\AA, and contains three different Al lines ($\lambda$ $\sim$ 16719.0, 16750.6 and 16763.4~\AA). The strongest Al~{\sc i} line is at $\lambda_{\rm Al}=16750.6$~\AA\ and is present in objects with spectral types between approximately K3 and M6-7. The line depth of Al~{\sc i} remains unchanged for the late-type K and early-type M stars, but then decreases at lower temperatures. The second 
feature, which is located around $\lambda_{\rm peak}=16745.9$~\AA\ is a flux bump that rises at lower temperatures. The peak flux is the result of an atmospheric transmission window (the absence of absorbers) in the star, and is coincident with the disappearance of Fe in the OH region. 
This flux peak is sensitive to \teff\ beginning in M4 stars and later. 
The contrary dependence of the Al and peak flux line depths to \teff\ is as useful at deriving \teff\ as the OH and Fe line depths, but at lower temperatures.

\subsection{Determining \teff}\label{sect:deter}

At the IGRINS spectral resolution the Fe line that we used is un-blended, the OH lines are blended but approximately equal in depth, the flux peak is created by the absence of absorption 
within the broad absorption defining the pseudo-continuum, and the broad Al line is blended with OH and CO at high and low temperatures, respectively. These characteristics of the lines make equivalent width  measurements inconsistent across a broad sampling of spectral types. Yet, we find that line-depths consistently trace \teff\ (see Figure~\ref{fig:ld}) and here we describe our methods.  

\subsubsection{Line-depth Measurements}\label{sect:ldepths}

\begin{figure*}
\centering
\includegraphics[width=0.9\textwidth]{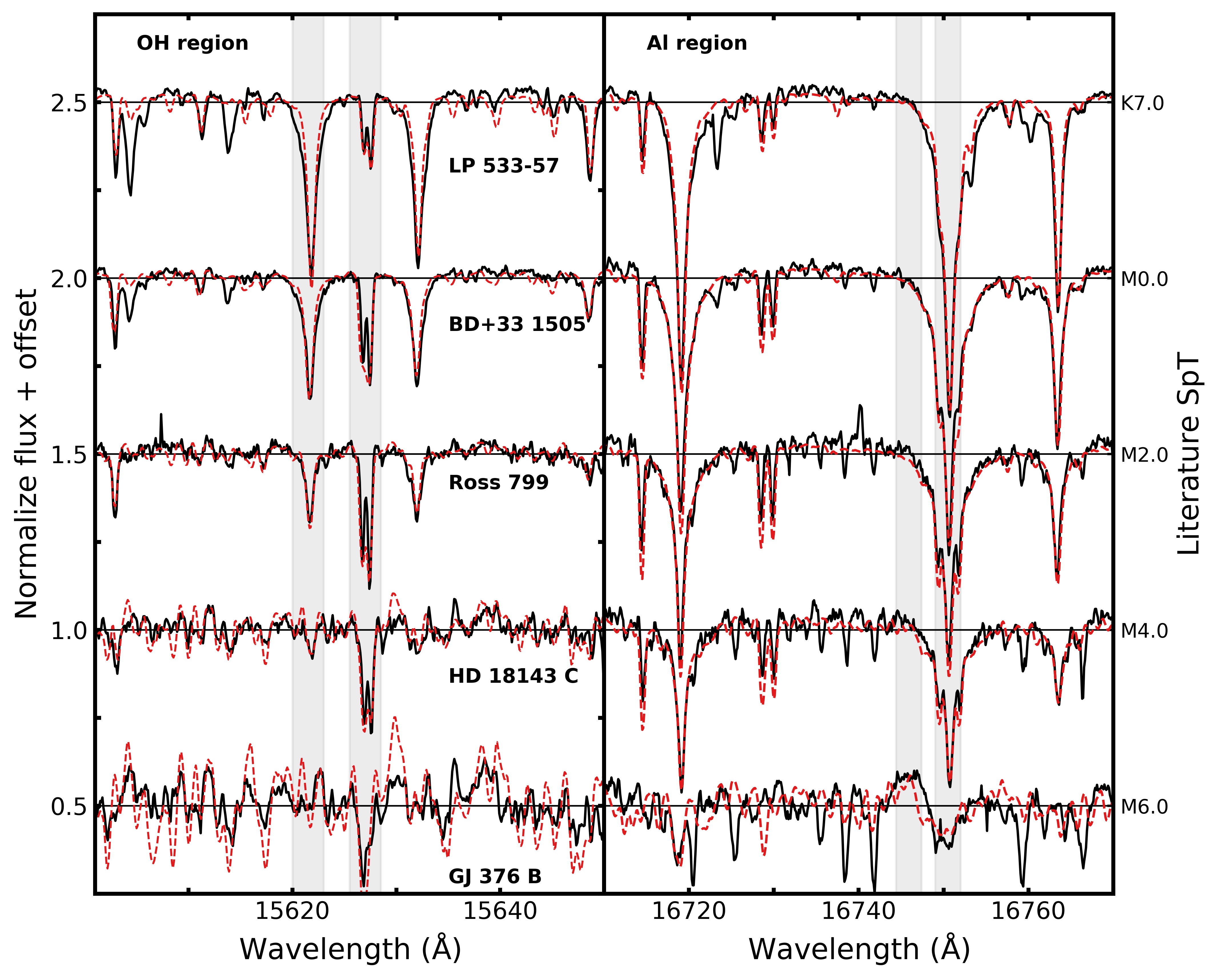}
\caption{A representative sample of IGRINS (black solid line) and synthetic BT-Settl (red dashed line) spectra around the OH region and the central 60~\AA\ of the Al region as a function of spectral classification. The horizontal lines represent the median flux across the region of interest and the level where line-depths are measured. These regions also include several other atomic (Fe~{\sc i}, Ti~{\sc i} and Ni~{\sc i}) and molecular (FeH, CN and CO) lines that are not labeled here and absent in the synthetic spectra.\label{fig:conti}}
\end{figure*}

As mentioned before, molecular lines dominate the atmospheres of cool stars and complicate the determination of a continuum level, which leads to inconsistencies in spectral normalization.
To address this issue, we computed the median flux across the entire OH or Al region and used this value to normalize our spectra within those regions. 
Continuum fitting using the average flux across the region, or a smoothed spectrum, did not produce a consistent definition of the continuum for all spectra. More complicated determinations of the continuum using iterative sigma clipping, or the upper quartile of the flux within the region, produces the same results as using the median but with some constant offset. 
The spectra in Figure~\ref{fig:conti} have been normalized by the median flux, and while this may not provide the most accurate determination of the continuum level, it produces repeatable measurements when applied uniformly across the entire analysis.
Since we normalize the BT-Settl models in the same fashion, and these models accurately reproduce the spectra of our stars, any inaccuracy in the normalization is consistently applied to all spectra. 

After normalizing the spectrum, we searched within $\pm$1.5~\AA\ of the the central wavelengths ($\lambda_{\rm OH}$, $\lambda_{\rm Fe}$, $\lambda_{\rm Al}$, $\lambda_{\rm peak}$) for the 
minimum flux value of the Fe~{\sc i}, OH, Al~{\sc i} lines and the maximum for the flux peak. We then computed the average flux ($\bar{f_\lambda}$) and the standard deviation of the mean ($\sigma_f$) within a window of 5~pixels ($\sim$1.5 resolution elements), centered at the min/max found previously. The measured line depth (flux peak height) is $d$ = 1 -$\bar{f_\lambda}$ and we assign $\sigma_f$ as the uncertainty.
The line depths were determined the same way in the observed and synthetic spectra. The line depths from the synthetic spectra defined a matrix of values for each spectral region, where the corresponding depths are identified by the unique \teff\ and $v \sin i$ combination of the model grid. 

\begin{figure*}
\centering
\includegraphics[width=0.8\textwidth]{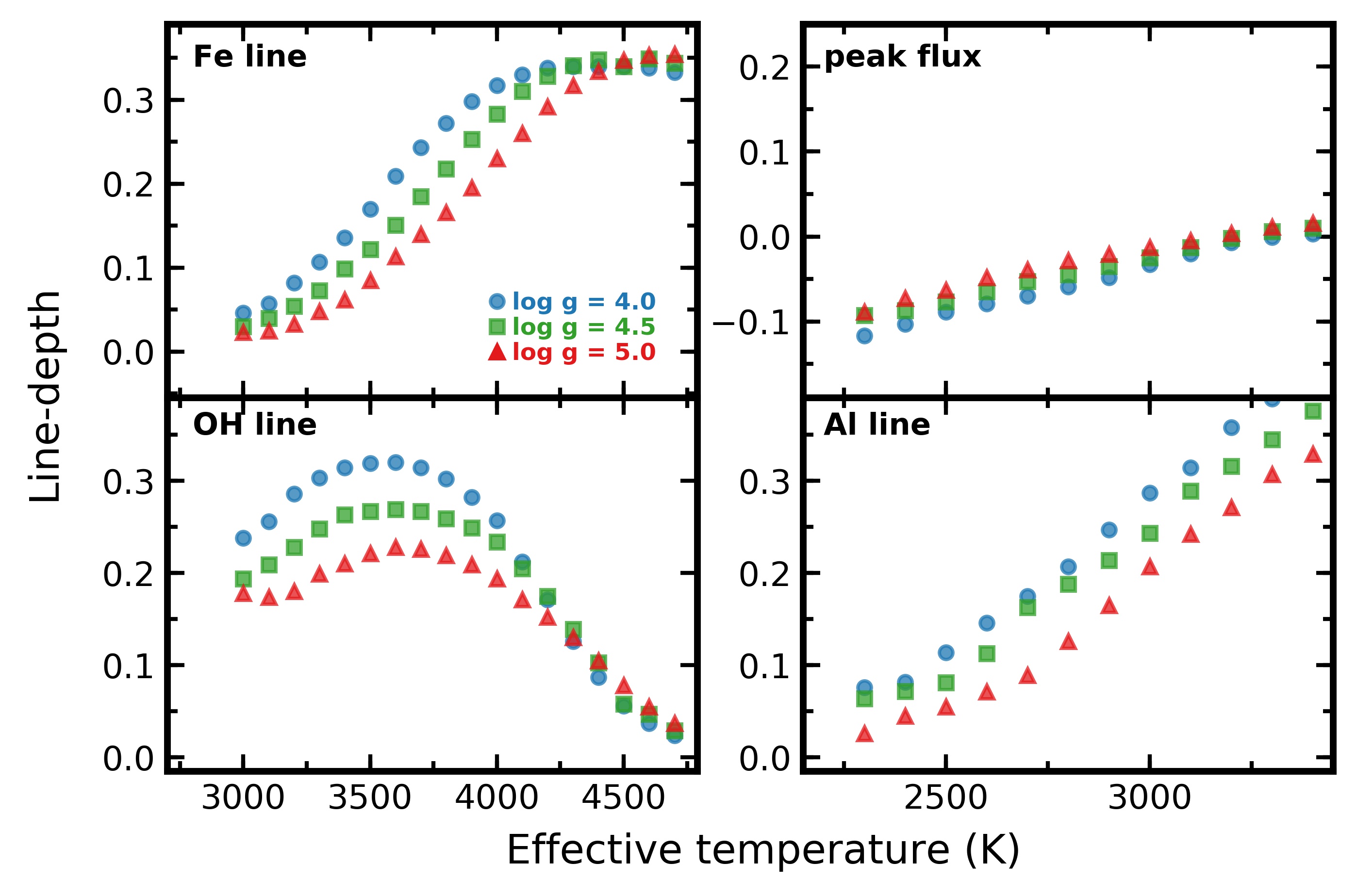}
\caption{BT-Settl synthetic line depths as a function of effective temperature, color-coded by surface gravity. The synthetic line-depths of both regions exhibit a strong contrary dependence to \teff\ and a weaker one to \logg.\label{fig:ld}}
\end{figure*}

In Figure~\ref{fig:ld} we present the synthetic line-depths as a function of \teff\ and \logg.  Adding the dependence to \logg\ in this figure help us to examine, in a qualitative way, how our \teff\ determination is modulated by variations in \logg. In the synthetic spectra, the Fe~{\sc i} line depth increases monotonically for \teff~$>3000$~K (SpT earlier than $\sim
$M5)  and it appears to saturate around 4500~K ($\sim$K4). By contrast, the OH depth increases more slowly to a maximum value at $\sim$3600~K ($\sim$M2) and then decreases up to  $\sim$4700~K ($\sim$K3). 

The synthetic line depth of the flux peak, which by our definition is negative since it is above the pseudo-continuum, increases monotonically from 2300~K ($\sim$M9) to $\sim$3200~K ($\sim$M4). 
The Al~{\sc i} line depth decreases linearly with decreasing \teff. 
The role of \logg\ in the Al region seems less important than in the OH region, since the changes on \teff\ produced by $\pm$0.5 in \logg\ are $\sim$120~K and $\sim$90~K for the flux peak and the Al~{\sc i} line. Additionally, increasing gravity reduces both the amplitude of the peak flux and the line depth of the Al line.

Another advantageous feature of these spectral regions is the range of SpT over which they are sensitive to \teff. Together, they allow us to determine the \teff\ scale for $\sim$K8 to M5 stars. While the Al region is appropriate for late-type objects ($\sim$M4 and later) the OH region is useful for SpTs earlier than $\sim$M5, having an overlapping zone of about 1 sub-class in SpT.
   
In the analysis that we present here we assume that \logg\ =~4.5 for all the targets in our sample and we adopted solar metallicity. These assumptions were made because our targets are nearby field stars that most probably reside in the thin disk \cite[eg. ][]{reyle02}. We investigate the impact of these assumptions on our final determination of \teff\ in Sections~\ref{sect:meta} and \ref{sect:gra}.

\subsubsection{A Precise \teff\ Sequence}\label{sect:temp_mean}
For each pair of line-depths we performed a $\chi^2$ minimization between the observed star and the synthetic line-depth grid corresponding to the star's $v \sin i$. The derived line-depth temperature (T$_{\rm LD}$) was taken as the weighted mean of the temperatures corresponding to the minimum and the two closest $\chi^2$ values. The uncertainty in the temperature determination ($\sigma_{T_{\rm LD}}$) was measured as:

\begin{equation}
\sigma_{T_{\rm LD}}  = \frac{n}{(n-1) W^2}\sum_i w^2(T_{i} - T_{\rm LD})^2
\end{equation}
where $n$(=3) is the number of measurements used in the weighted average,  $W = \sum_i w$,  $w$ is the weight ($= 1/ \chi^{2}$), $T_i$ is the model temperature and $T_{\rm LD}$ is the weighted mean temperature. When the minimum $\chi^2$ corresponded to the lower or upper edges of the synthetic line-depth grid then the derived temperature was given a null value. 

From the measured line depths of the K and M stars in our sample we assigned \teff\ to each star based on the line depths of the synthetic spectra.
\teff\ was determined by means of the OH region in 116 stars, the Al region in 92 stars, and using both spectral regions for 46 stars.  

While the IGRINS spectra were well matched by the BT-Settl models (Figure~\ref{fig:conti}), the temperature scale obtained using theoretical grids are generally precise, but also inaccurate. The inaccuracy stems from the different physical assumptions of the stellar structure, atomic and molecular line lists, and the modeler's treatment of the line strengths. 

In the following section we used the empirical color-Temperature relation of \citet{mann15} to take into account discrepancies in the temperature scale between models and observation.

\subsubsection{Accurate \teff's for K and M Stars}\label{sect:cal}
\cite{mann15} used accurate spectrophotometric calibrations to determine \teff, bolometric flux, metallicity, and stellar radii for 183 nearby K7 -- M7 stars. Those \teff\ values were calibrated by means of temperatures determined from interferometric data for 29 stars, resulting in an empirical temperature scale. 

Interferometrically determined temperatures are accurate for the range of stellar parameters that are covered by the sample itself. 
For the 51 stars we have in common with the Mann et al. sample, only 14 of these have interferometric data.
We chose to calibrate our line-depth temperatures from the models above to empirical scale, by means of their (r - J) color-Temperature relation, instead of using the stars in common. The (r - J) color-Temperature relation determined by \cite{mann15} is tied to the interferometric stars and is valid for 2700~$<$~\teff~$<$~4100~K:

\begin{equation}
T_{\rm emp} =3500 \times (a + bX + cX^2 + dX^3 + eX^4)
\end{equation}
where  $a$, $b$, $c$, $d$, and $e$ are the polynomial coefficients found by \cite{mann15}, with values of 2.84, -1.3453, 0.3906, -0.0546 and 0.002913, respectively,  and $X$ is the (r - J) color in magnitudes. We retrieved the available r- and J-band photometry for all our sample from the AAVSO All-Sky Photometric Survey \cite[APASS;][]{henden12} and the Two Micron All Sky Survey \cite[2MASS;][]{cutri03}, and then computed empirical temperatures using the above equation. 
The photometric data and the (r - J) color temperatures are reported in Table~\ref{tab:teffs}. 
The calibration sample comprises 106 stars in the Mann et al. sample and the IGRINS sample, from which 66 were determined with the Al region and 64 from the OH region (24 stars are in both regions).

\begin{figure*}
\centering
\includegraphics[width=\textwidth]{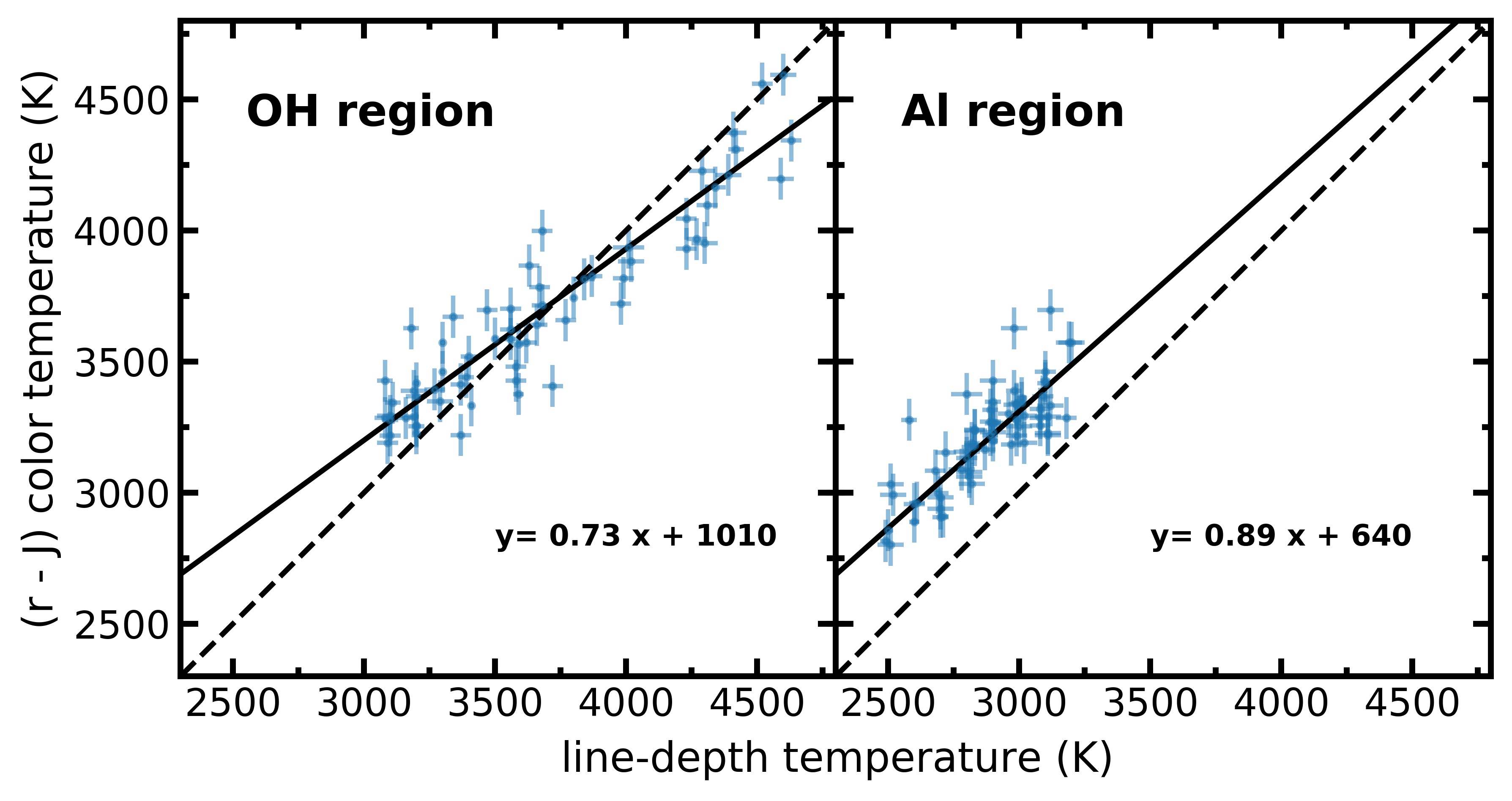}
\caption{Comparison between the \teff\ determined in this work through Al and OH region and their corresponding empirical temperatures.The dashed line represents the one-to-one relation while the solid one is the weighted linear fit. Errors on empirical temperatures are all assumed to be 80~K, which is the quadrature sum of the typical spectroscopic error (60~K) and the dispersion of the calibration (58~K) as reported in \citet{mann15}.\label{fig:c_both}}
\end{figure*}

Figure~\ref{fig:c_both} illustrates how the derived temperatures from both spectral regions correlate linearly with the (r - J) color temperatures. 
The temperatures determined using the Al region primarily exhibit an offset of $\sim$640~K with respect to the empirical temperatures, while those determined using the OH region display a steeper slope with respect to their empirical counterparts.  
The equations in Figure~\ref{fig:c_both} were used to convert the precise line-depth temperature sequence into an accurate one calibrated against the Mann et al. sample. 
These empirically calibrated temperatures ($T_{\rm spec}$) are considered the final measurements.  
We assigned for the stars with temperatures determined in both the OH and Al regions an average of their corresponding calibrated temperatures, and the sum in quadrature of the individual errors is the final uncertainty.

\begin{deluxetable*}{lrrrrrrrrrrrr}
\tabletypesize{\tiny}
\tablecaption{Basic information as well as our results for the  first 20 entries of our sample. We compile SpT, r and J magnitudes, the empirical temperatures (T(r-J)), rotational velocity, the four line-depths, an identification number corresponding to the source of the temperature being, 1 from OH region, 2 from Al region, 3 from the average of both regions and 4 if it is a limit, and in the final column  we report our $T_{\rm spec}$. The full version of this table will be available in the online version of the paper.\label{tab:teffs}}
\tablecolumns{13}
\tablewidth{0pt}
\tablehead{
\colhead{Star}& \colhead{SpT}& \colhead{Ref.\tablenotemark{a}}&\colhead{J}& \colhead{r}& \colhead{T(r-J)} & \colhead{$v \sin i$}& \multicolumn{4}{c}{normalize flux line-depths}&\colhead{reg} &\colhead{$T_{\rm spec}$\tablenotemark{b}} \\
\colhead{ }& \colhead{ }& \colhead{}& \colhead{(mag) }& \colhead{(mag)}&\colhead{(K)}& \colhead{(km s$^{-1}$)}& \colhead{Fe~{\sc i}}& \colhead{OH}& \colhead{peak}& \colhead{Al~{\sc i}} &\colhead{} &\colhead{(K)} }
\startdata
LP 699-32	&	0	&	1	&	10.67	&	15.59	&	2889	&	10	&		--		&		--		&	-0.075	$\pm$	0.006	&	0.127	$\pm$	0.006	&	2	&	2950	$\pm$	110	\\
NLTT 55442	&	0	&	1	&	10.39	&	15.04	&	2962	&	18	&		--		&		--		&	-0.069	$\pm$	0.004	&	0.121	$\pm$	0.004	&	2	&	2960	$\pm$	110	\\
LSPM J2206+4322W	&	0	&	1	&	10.78	&	--	&	--	&	--	&	0.058	$\pm$	0.002	&	0.190	$\pm$	0.020	&	-0.044	$\pm$	0.005	&	0.320	$\pm$	0.010	&	3	&	3360	$\pm$	90	\\
G 194-18	&	0	&	1	&	10.56	&	13.74	&	3427	&	$<$7	&	0.041	$\pm$	0.003	&	0.200	$\pm$	0.100	&	-0.031	$\pm$	0.004	&	0.308	$\pm$	0.006	&	3	&	3330	$\pm$	90	\\
G 122-46	&	0	&	1	&	10.59	&	--	&	--	&	8	&	0.040	$\pm$	0.010	&	0.250	$\pm$	0.020	&		--		&		--		&	1	&	3270	$\pm$	140	\\
NLTT 19346	&	0	&	1	&	11.76	&	--	&	--	&	--	&		--		&		--		&	-0.073	$\pm$	0.004	&	0.116	$\pm$	0.003	&	2	&	2950	$\pm$	110	\\
UCAC4 368-064862	&	0	&	1	&	9.27	&	11.81	&	3702	&	$<$7	&	0.160	$\pm$	0.007	&	0.240	$\pm$	0.020	&		--		&		--		&	1	&	3610	$\pm$	140	\\
$[{\rm RSP2011}]$ 315	&	0	&	1	&	11.01	&	14.23	&	3413	&	--	&	0.110	$\pm$	0.010	&	0.220	$\pm$	0.020	&		--		&		--		&	1	&	3470	$\pm$	140	\\
UCAC4 445-057351	&	0	&	1	&	9.76	&	13.25	&	3320	&	12	&		--		&		--		&	-0.042	$\pm$	0.003	&	0.265	$\pm$	0.003	&	2	&	3380	$\pm$	120	\\
LP 611-70	&	0	&	1	&	9.51	&	--	&	--	&	9	&	0.235	$\pm$	0.009	&	0.210	$\pm$	0.030	&		--		&		--		&	1	&	3790	$\pm$	130	\\
G 43-43	&	0	&	1	&	9.41	&	12.11	&	3623	&	$<$7	&	0.160	$\pm$	0.004	&	0.240	$\pm$	0.020	&		--		&		--		&	1	&	3610	$\pm$	140	\\
UCAC4 545-148763	&	0	&	1	&	9.17	&	11.50	&	3826	&	8	&	0.255	$\pm$	0.006	&	0.200	$\pm$	0.020	&		--		&		--		&	1	&	3840	$\pm$	140	\\
2MASS J12371238-4021480	&	0	&	1	&	9.47	&	12.88	&	3347	&	--	&		--		&		--		&	-0.033	$\pm$	0.001	&	0.218	$\pm$	0.002	&	2	&	3220	$\pm$	110	\\
2MASS J04435750+3723031	&	0	&	1	&	12.22	&	--	&	--	&	--	&		--		&		--		&	-0.083	$\pm$	0.004	&	0.113	$\pm$	0.003	&	2	&	2950	$\pm$	110	\\
BD+45 598	&	K0.0	&	2	&	7.62	&	8.80	&	--	&	19	&	0.247	$\pm$	0.004	&	0.000	$\pm$	0.002	&		--		&		--		&	4	&	4440	$\pm$	130	\\
HD 285690	&	K0.0	&	2	&	7.88	&	9.24	&	--	&	10	&	0.400	$\pm$	0.010	&	0.008	$\pm$	0.003	&		--		&		--		&	4	&	4440	$\pm$	130	\\
HD 286363	&	K0.0	&	3	&	8.18	&	9.72	&	--	&	11	&	0.400	$\pm$	0.010	&	0.025	$\pm$	0.003	&		--		&		--		&	4	&	4440	$\pm$	130	\\
HD 285482	&	K0.0	&	3	&	8.11	&	9.56	&	--	&	11	&	0.400	$\pm$	0.010	&	0.016	$\pm$	0.001	&		--		&		--		&	4	&	4440	$\pm$	130	\\
HD 285876	&	K0.0	&	4	&	8.67	&	10.51	&	4212	&	11	&	0.380	$\pm$	0.010	&	0.110	$\pm$	0.010	&		--		&		--		&	1	&	4210	$\pm$	140	\\
\enddata
\tablenotetext{a}{Reference for SpT shown in SIMBAD at the time of the query (March 2019).}
\tablenotetext{b}{The error reported is just the random uncertainties, while the systematic ones were estimate in Section~\ref{sect:err_fin} and are of $\pm$120~K.}
\tablereferences{(1)No specified;
(2)\citet{2007MNRAS.374..664C}; 
(3)\citet{1995AAS..110..367N}; 
(4)\citet{2014AJ....148..108B}
(5)\citet{1989ApJS...71..245K};  
(6)\citet{1975mcts.book.....H};  
(7)\citet{1999MSS...C05....0H};
(8)\citet{2007AJ....133.2524W};
(9)\citet{1986AJ.....91..144S};   
(10)\citet{1986AJ.....92..139S};  
(11)\citet{1985ApJS...59..197B};  
(12)\citet{2004AJ....128..463R};
(13)\citet{2010MNRAS.403.1949K};  
(14)\citet{1993ApJ...419L..89F};  
(15)\citet{2006AA...460..695T};  
(16)\citet{2015AA...577A.128A};
(17)\citet{2003AJ....126.2048G};  
(18)\citet{2002AJ....123.2002H};  
(19)\citet{2012AJ....143...80S};  
(20)\citet{kirkpatrick91};
(21)\citet{2013AJ....145..102L};
(22)\citet{2006AJ....132..866R};  
(23)\citet{2009ApJ...699..649S};  
(24)\citet{2009AA...504..981B};
(25)\citet{2012AJ....143..114S};  
(26)\citet{mann13};  
(27)\citet{2014AJ....147..146K};  
(28)\citet{1974ApJS...28....1J};
(29)\citet{2014AJ....147...20N};  
(30)\citet{1997PASP..109..643P};  
(31)\citet{2014MNRAS.438.2413V}; 
(32)\citet{2015ApJ...802L..10T};
(33)\citet{2015AJ....149..106D};  
(34)\citet{2007AJ....133.2825R};  
(35)\citet{1983SAAOC...7..106W};  
(36)\citet{2013MNRAS.431.2745G};
(37)\citet{2014AJ....148...36A};  
(38)\citet{rojas-ayala12};  
(39)\citet{2006PASP..118..671R};  
(40)\citet{2008MNRAS.384..150L};
(41)\citet{mann16};  
(42)\citet{herczeg14};  
(43)\citet{2006AA...460L..19M};  
(44)\citet{mann14};
(45)\citet{2005AA...442..211S};  
(46)\citet{2015ApJS..216....7B};  
(47)\citet{2015ApJS..219...33G};  
(48)\citet{2015ApJ...812....3W};
(49)\citet{2012MNRAS.419.3346G};  
(50)\citet{2007AJ....133.2258S}.}

\end{deluxetable*}

\section{Results and discussion}\label{sect:resul}
\begin{figure}
\centering
\includegraphics[width=\columnwidth]{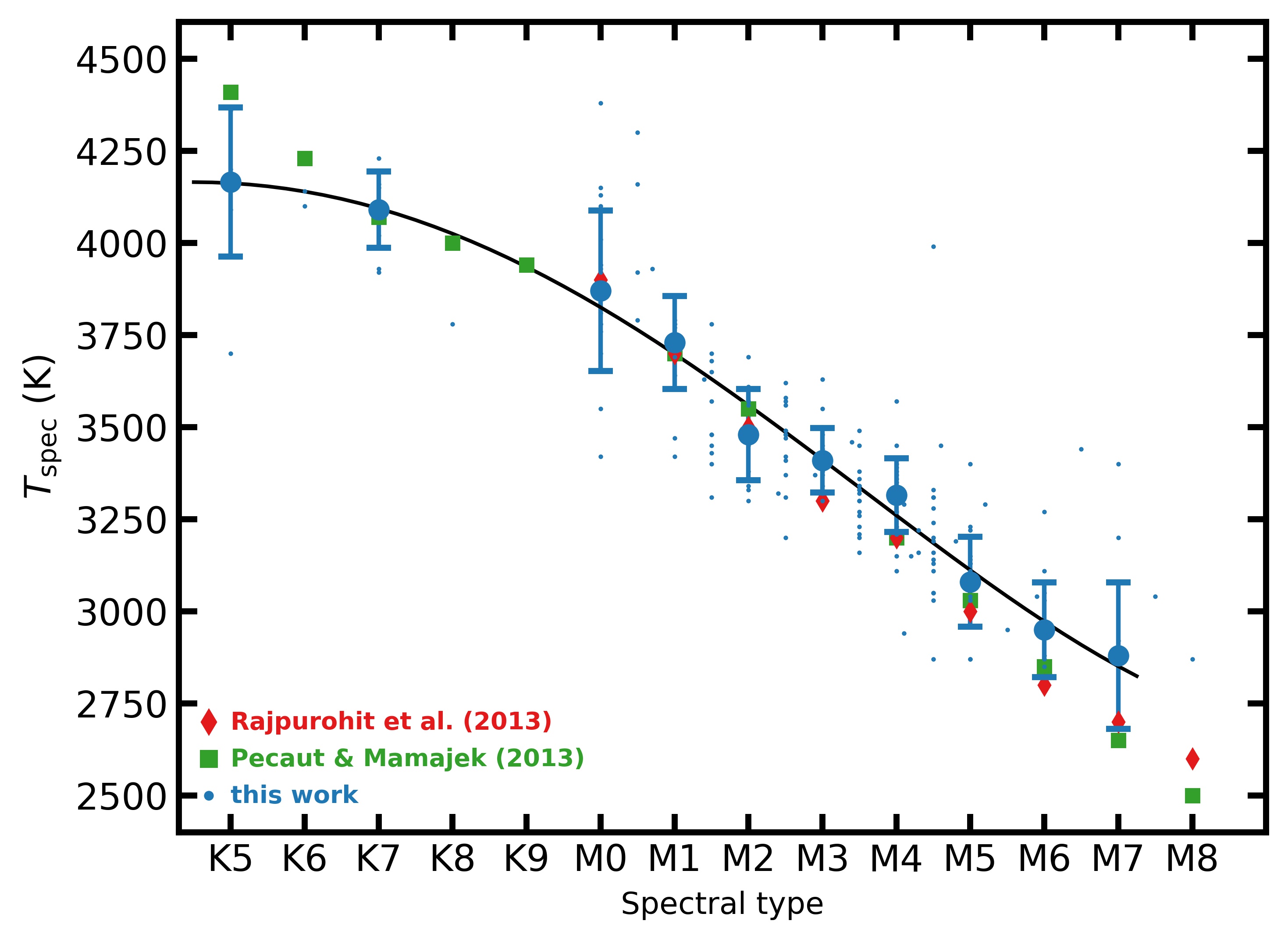}
\caption{Our $T_{\rm spec}$ (small circles) as a function of the literature SpT. The solid line is a weighted fourth degree polynomial fit to the median values of $T_{\rm spec}$ (large circles) to non-fractional SpT with more than two stars, while the error bars represent the one standard deviation level. The squares are the temperature scale for dwarfs stars of \citet{pecaut13} and the diamonds are the results of \citet{rajpurohit13}.\label{fig:scale}}
\end{figure}

In Table~\ref{tab:teffs} we report the temperatures we derive along with basic information for all 254 K and M field stars. 
Although the compiled SpT of our sample is precise to only $\pm$1-2 subtypes, we constructed a 
SpT--$T_{\rm spec}$ relation (see Figure~\ref{fig:scale} and Table~\ref{tab:spttef}) to compare with temperature scales determined for dwarf stars by \cite{pecaut13} and the median results of \cite{rajpurohit13}. Both studies used BT-Settl models, but with the solar composition of \cite{asplund09} and \cite{caffau11}, respectively. \cite{pecaut13} determined \teff\ by using the Spectral Energy Distribution Fitting method (SEDF; \citealp{masana06}), which simultaneously fits the observed and synthetic photometry. On the other hand, \cite{rajpurohit13} compared low- and medium-resolution ($\Delta\lambda$~=~10 and 4~\AA) optical spectra with BT-Settl models to determine temperature.

We found good agreement between \cite{rajpurohit13}, \cite{pecaut13} and this work for objects with SpT between K6--M6, where the maximum difference with our median temperatures is 150~K, being of the order of our typical error ($\sigma_{\rm typ}$= 140~K). In the cases of the K5 and M7 bins these differences increased up to 245~K ($\sim$1.8$\sigma_{\rm typ}$). A fourth degree polynomial fit to the median temperature per SpT bin provides the equation:
\begin{equation}
T_{spec} = a + bX + cX^2 + dX^3 + eX^4
\end{equation}
where $X$ is the SpT and takes numerical values between 4 and 17 (equivalent to SpT K4 to M7) and $a, b, c, d,$ and $e$ are the fitted polynomial coefficients equal to 3973.570, 74.705, -4.140, -0.821, 0.034, respectively.

\begin{deluxetable}{lcccc}
\tabletypesize{\normalsize}
\tablecaption{Median \teff\ and standard deviation, along with the temperature determined by PM13 = \citet{pecaut13} and R13 = \citet{rajpurohit13} for each SpT. Stars with intermediate spectral classifications were not include.\label{tab:spttef}}
\tablecolumns{5}
\tablewidth{0pt}
\tablehead{
\colhead{SpT} & \colhead{\# stars} & \colhead{\teff~$\pm$~$\sigma$} & \colhead{PM13} &\colhead{R13} \\ 
\colhead{} & \colhead{} & \colhead{(K)} & \colhead{(K)} & \colhead{(K)}} 
\startdata
K5 & 6   & 4165 $\pm$ 200  & 4410 & --\\
K6 & 2   & 4120 $\pm$ 40\tablenotemark{a}   & 4230 & --\\
K7 & 8   & 4090 $\pm$ 100  & 4070 & --\\
M0 & 19  & 3870 $\pm$ 220 & 3870 & 3900 \\
M1 & 10  & 3730 $\pm$ 130 & 3700 & 3700 \\
M2 & 11  & 3480 $\pm$ 120 & 3550 & 3500\\
M3 & 14  & 3410 $\pm$  90 & 3410 & 3300\\
M4 & 28  & 3315 $\pm$ 100 & 3200 & 3200\\
M5 & 19  & 3080 $\pm$ 120 & 3030 & 3000\\
M6 &  9  & 2950 $\pm$ 130 & 2850 & 2800\\
M7 &  7  & 2880 $\pm$ 200 & 2650 & 2700
\enddata
\tablenotetext{a}{The average temperature and the difference between individual determinations is reported.}
\end{deluxetable}

\subsection{Sources of uncertainty in $T_{\rm spec}$ measurements}

Besides the literature SpT, another possible source of scatter in Figure~\ref{fig:scale} could be the fixed values we chose for \logg\ and [Fe/H]. 
Although these are reasonable assumptions for our sample of field dwarfs, in the next sections we investigate the potential impact of these two parameters on the \teff\ scale.

\subsubsection{Metallicity effects}\label{sect:meta}

Along with \teff, \cite{mann15} also reported metallicities determined from equivalent widths of atomic features in low-resolution near-infrared spectra. Such metallicities were calibrated by means of wide binary systems with FGK primary stars and M dwarf companions \citep{mann13,mann14}.

For the 51 stars in common with \cite{mann15}, we explore trends related to [Fe/H]. Although this comparison is not independent, since we corrected our temperatures with the \cite{mann15} color-Temperature relation, it is still meaningful to better understand the role of [Fe/H] on the derived temperature scale.

\begin{figure*}
\centering
\includegraphics[width=\textwidth]{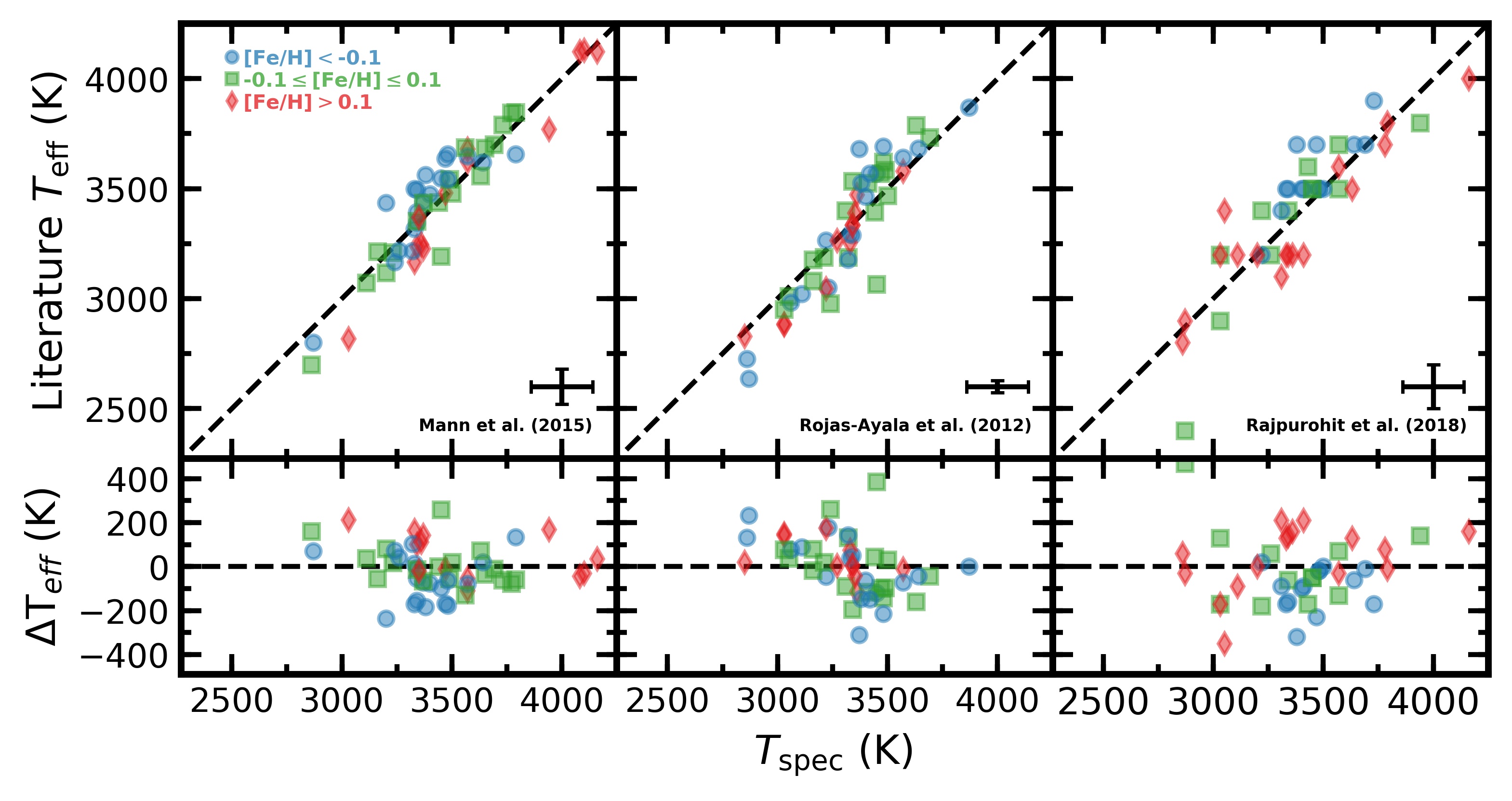}
\caption{Comparison between our $T_{\rm spec}$ (x-axis) and those determined by \citet{mann15} (left), \citet{rojas-ayala12} (middle)  and \citet{rajpurohit13} (right), color-coded by their metallicity determinations. The lower panels show the residuals of our derived \teff\ minus the literature temperatures. The mean error in each panel is about 80, 30 and 100~K, for \citet{mann15},\citet{rojas-ayala12} and \citet{rajpurohit13}, respectively, while our typical error is $\sim$140~K. See text for discussion.
\label{fig:c_meta}}
\end{figure*}

The left panel of Figure~\ref{fig:c_meta} depicts the comparison between $T_{\rm spec}$ and the Mann et al. temperatures, color-coded by the metallicities of Mann et al. The [Fe/H] of 
the stars in common with \cite{mann15} spans from -0.38 to +0.39 dex, which we have classified into three categories: metal-poor ( [Fe/H] $<$ -0.10), solar composition (-0.10 $\leq$ [Fe/H] $\leq$ +0.10) and metal-rich ([Fe/H] $>$ +0.10) stars. Using these metallicity classifications we identified 20 metal-poor, 18 solar composition and 13 metal-rich stars. We computed the reduced chi-square ($\chi_\nu^2$) between our observations and the literature values as a measurement of the agreement between the two temperatures. We found $\chi_\nu^2$ value of 1.2, 0.8 and 1.2 for the metal-poor, solar composition and metal-rich stars. 

The good agreement with the solar composition stars is not surprising since we derived \teff\ from solar metallicity models. We expected some temperature variations as the stellar metallicity departs from the solar value because the line-depths appear deeper/shallower as [Fe/H] increases/decreases.

\cite{rojas-ayala12} also estimated \teff\ and [Fe/H], but using equivalent widths of the Ca ($\sim$22,050~\AA) and Na ($\sim$22,630~\AA) lines, as well as the H$_2$O-K2 index, in low-resolution (R$\sim$2,700) K-band spectra. 
We have in common with \cite{rojas-ayala12} 47 stars that we compare in the middle panel of Figure~\ref{fig:c_meta}. The overall $\chi_\nu^2$ of this comparison is 1.7 and we found a slight trend, which highlights a 
systematic difference between our methods since \teff\ smaller (larger) than $\sim$3300~K seems to be underestimated (overestimated). 
Such a trend was also pointed out by \cite{mann15}. We found $\chi_\nu^2=2.0$ for 18 metal-poor, $\chi_\nu^2=2.2$ for 18 solar composition and $\chi_\nu^2=0.7$ for 11 metal-rich stars. The cause of the trend in $\Delta$\teff\ compared to \cite{rojas-ayala12} is likely because they used an older version of the BT-Settl models, which employs the solar abundances of \cite{asplund09} rather than the \cite{caffau11} and an older versions of line lists. 

Finally, in the right panel of Figure~\ref{fig:c_meta}, we compared our determinations with those made by \cite{rajpurohit18b}, which determined \teff, \logg\ and [Fe/H] from optical and near infrared ($\sim$7,500--17,000~\AA) high-resolution ($R$=90,000) spectra. The general comparison resulted in $\chi_\nu^2=1.7$, while the comparison by category is $\chi_\nu^2=2.0$, $\chi_\nu^2=1.1$, and $\chi_\nu^2=1.9$, for 14 metal-poor, 11 solar composition, and 16 metal-rich stars, respectively. There is not an obvious trend with metallicity, but the comparison shows a larger dispersion than our comparison to \citet{mann15}. 

The result of these three comparisons reveals that our temperature scale is consistent with previous ones, giving us the ability to determine accurate \teff\ for any star within the IGRINS archive without the necessity of extra data, such as photometry.  
Additionally, we found that a difference in metallicity of $\Delta$[Fe/H]~=~$\pm$0.4 will have the effect of change our temperatures by $\Delta T_{\rm spec}$~=~$\mp$100~K. In other words, our method will produce hotter and cooler temperatures for metal-rich and metal-poor stars, respectively.

Another important point comparison between this work and previous works is the value of \logg.
Since we calibrated our temperatures with the relationship from \cite{mann15}, 
the impact of using different \logg\ values was taken into account, as the good
agreement ($\chi_\nu^2=1.0$) showed. However, the differences found with \cite{rojas-ayala12} and \cite{rajpurohit18b} could be caused by \logg\ differences since they measured \logg\ rather than making it a fixed quantity.

\subsubsection{Surface gravity effects}\label{sect:gra}

To characterize the effects of surface gravity on our temperature sequence we chose synthetic models with \logg~=~4.0 and $5.0$ and determined \teff\ following the same line-depth method outlined in Section\ref{sect:deter}. 
With this approach we treat the synthetic spectra as a star with a known \logg\ value that we determine its temperature for with the \logg~=~4.5 models. 

\begin{figure}
\centering
\includegraphics[width=\columnwidth]{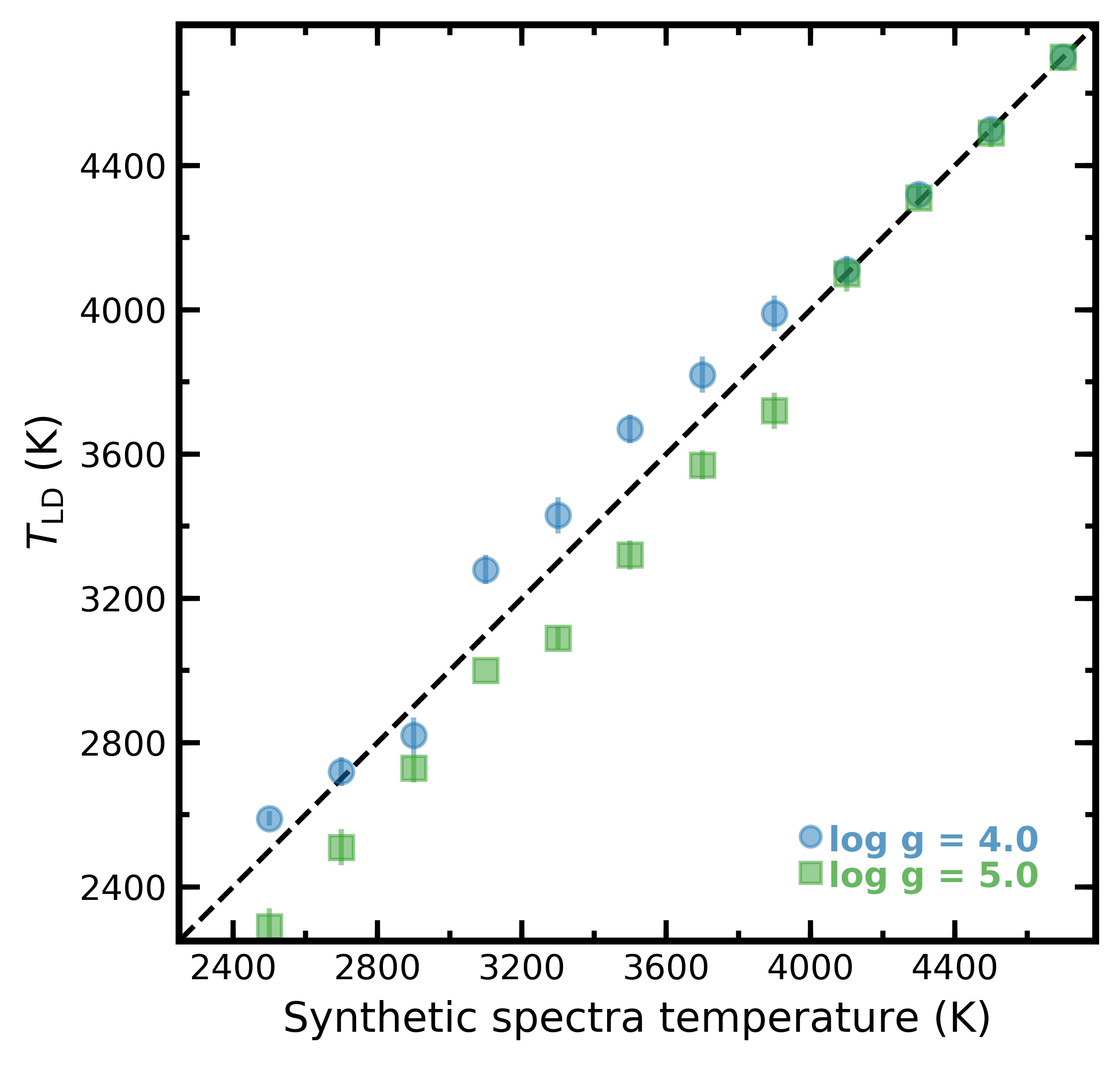} 
\caption{Line-depth temperature as function of synthetic spectra temperature with \logg\ =~4.0 (circles) and 5.0 (squares) dex. We determined the temperatures with a fixed \logg\ of 4.5 dex. For synthetic spectra with \teff\ $<$~3100~K, the determined temperature comes from the Al region, while the other from the OH region.\label{fig:gra_ef}}
\end{figure}

In Figure~\ref{fig:gra_ef} we show the results obtained in this test. We found that \logg\ is not important for \teff~$\gtrsim$~4100~K, an advantage of the OH region seen also in Figure~\ref{fig:ld}. 
For the synthetic spectra with \logg~=~4.0 and \teff\ between 3100 and 3900~K, we find hotter temperatures of $\sim$140~K on average.
For the synthetic spectra with \logg~=~5.0 our method recovered temperatures on average $\sim$160~K cooler. Below $\sim$3100~K the behavior is slightly different for \logg~=~4.0, in that the temperatures cross the one-to-one line, while for spectra with \logg~=~5.0 are consistently cooler. 

Averaging the mean differences found in the different temperature ranges we establish that a change in \logg\ of 0.5, will modify our \teff\ by $\sim$150~K. This effect will result in hotter temperatures for stars with \logg\ lower than 4.5 and viceversa. 
Stars with \logg\ $=4.5$ will show no systematic offset in \teff\ due to surface gravity assumptions.

\subsubsection{Rotational velocity effects}\label{sect:rota}
\begin{figure}
\centering
\includegraphics[width=\columnwidth]{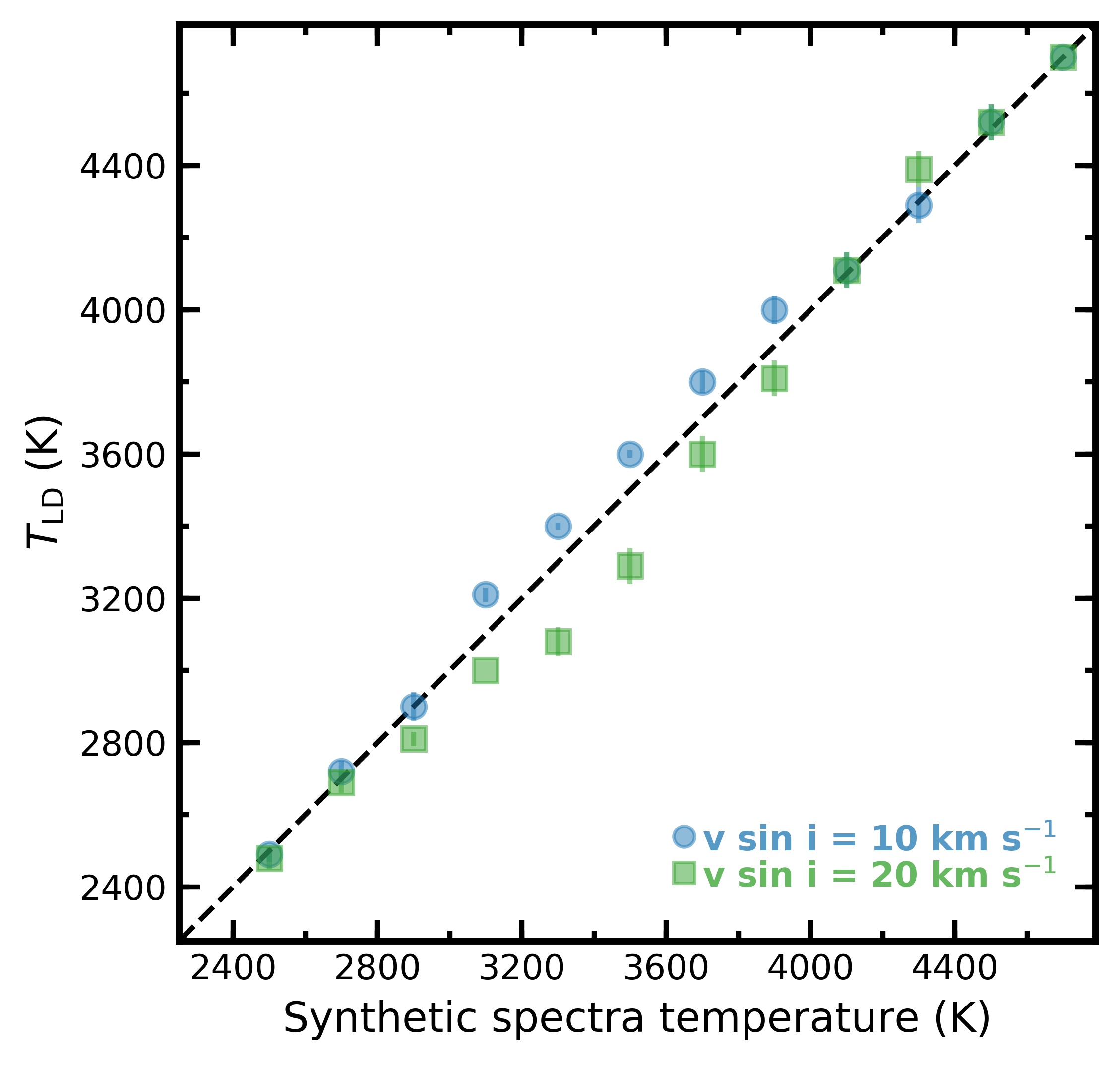}
\caption{Line-depth temperatures as function of the synthetic spectra temperature with $v \sin i$ =~10 (circles) and 20~km~s$^{-1}$ (squares). We determined the temperatures with a fixed $v\sin i$ of 15~km~s$^{-1}$. For synthetic spectra with temperatures greater than 3100~K, the determined temperature comes from the Al region, while the rest come from the OH region.\label{fig:ef_rota}}
\end{figure}

As in the last section, we used synthetic spectra with different $v\sin i$ values to assess the uncertainty introduce by an erroneous $v\sin i$. We tested $\Delta v\sin i $ = 5~km~s$^{-1}$ and our findings are shown in the Figure~\ref{fig:ef_rota}. For \teff~$<$~3000~K and \teff~$>$~4000~K the temperatures are less affected by a wrong $v\sin i$ value, with differences of the order of 20~K. The remaining temperatures appear cooler in average 130~K for fast rotators, while for slow rotators they are hotter by $\sim$100~K, therefore we consider that a difference in $v\sin i$ of $\pm$5~km~$s^{-1}$ from our calculated value has the effect of changing \teff\ up to 120~K. Such an effect will increase the derived temperature if $v\sin i$ is underestimated and viceversa. 
Stars with $v\sin i$ determined to within $\pm$2~km~$s^{-1}$ of the actual value, which is the case for much of our sample, will show minimal systematic offset in temperature due to $v\sin i$ errors.

\subsubsection{Cumulative Uncertainty Budget}\label{sect:err_fin}
In the case that we properly match the observed star's properties to our model grid (\logg~=~4.5, [Fe/H]~=~0.0, $v \sin i$~=~7~--~55~km~s$^{-1}$, spectral resolution of 45,000 and no $\alpha$-element enrichment) our $T_{\rm spec}$ uncertainties are driven by the calibration sample and are $\sim$140~K.
To have an estimate of the systematic error on our temperature determinations, we  
considered three different sources of error:  [Fe/H], \logg\ and $v\sin i$.
Linearly interpolating from the previous error analysis to the typical uncertainties for [Fe/H], $v\sin i$ and \logg\ in our sample (which are 0.25, 3~km~s$^{-1}$ and 0.25, respectively) we find a systematic uncertainty as high as 120~K.
For most of the stars in our sample, the errors presented in Table~\ref{tab:teffs} should properly account for calibration errors and small deviations from the model grid.
However, the systematic uncertainty of 120~K should be added for those objects with known outlier properties. 
From the examination of each contributing stellar parameter, temperature determinations can be further corrected for stars that have known properties that depart from the fixed values chosen in this study.


\subsection{Line Depth Ratios for Temperature Determination}
To support the broad applicability of our method, we obtained a mathematical expression that represents our temperature scale. We constructed a relation between \teff\ and the line-depth ratio (LDR) in each region. The LDR technique should be less sensitive to broadening processes that affect line-depths nearly equally, such as, resolution effects or veiling\footnote{The veiling is a continuum emission produce by the accretion of material onto the young star. This process reduces the 
depth of the photospheric lines.} in Young Stellar Objects (YSOs).  

\subsubsection{LDR vs. $T_{\rm spec}$}

\begin{figure*}
    \centering
    \includegraphics[width=0.9\textwidth]{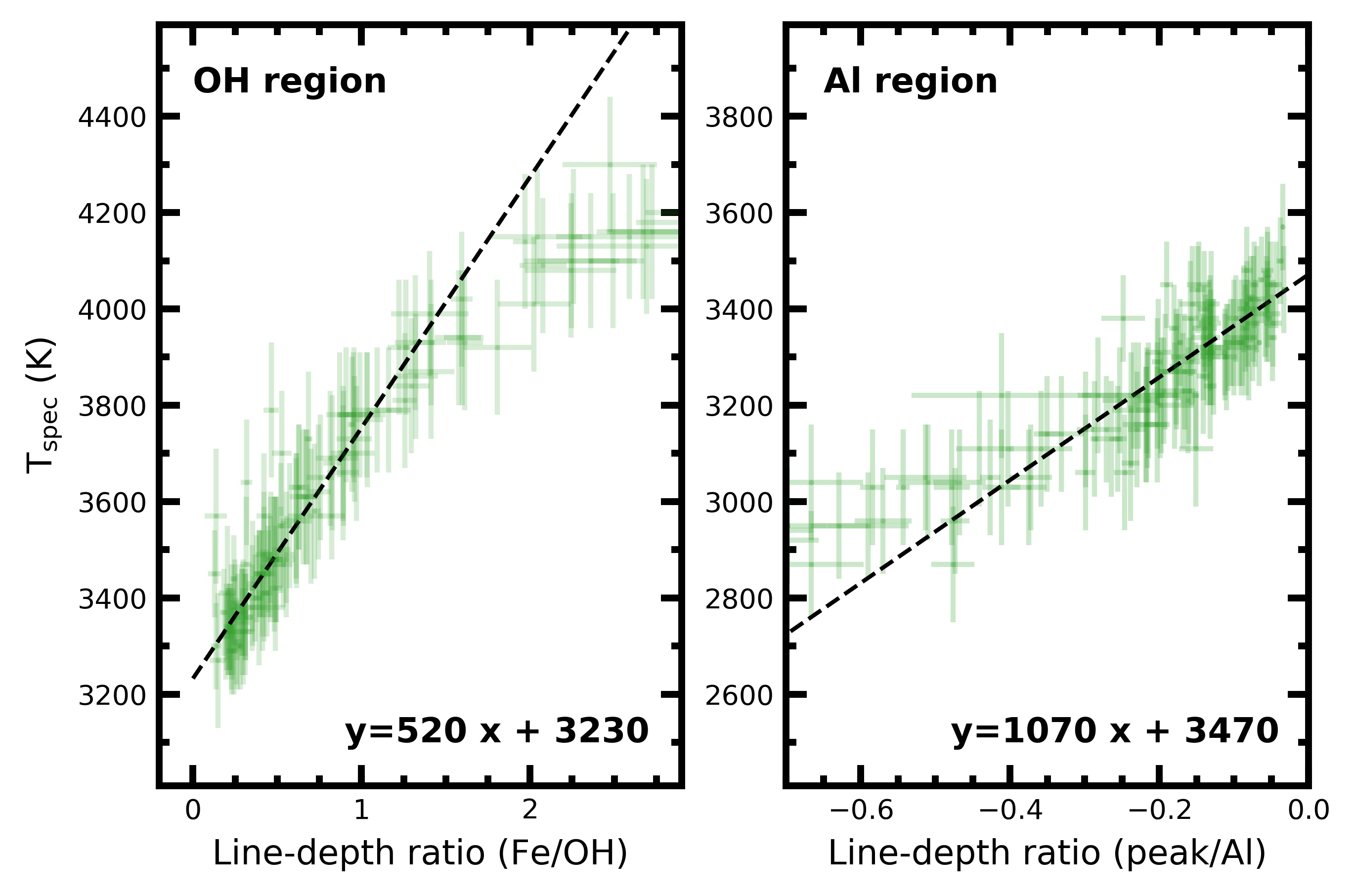}
    \caption{$T_{\rm spec}$ as function of LDR for both regions. The dashed line represents a linear fit to the data enclosed by LDR(Fe/OH)~$<$~1.5 and LDR(peak/Al)~$>$-0.5.\label{fig:ldr}}
\end{figure*}

In Figure~\ref{fig:ldr}, we show $T_{\rm spec}$ as a function of the ratio between the line-depth of the peak and the Al~{\sc i} (right panel) and between the Fe~{\sc i} and OH (left panel) lines. Such relations show, as expected, a good correlation between temperature and LDR within the range of 3000-4000~K. 
However, both LDRs exhibit a plateau at the hot and cool ends of the temperature sequence.
The plateau in the OH region is produced by the reduction of the flux in the OH line at $T_{\rm spec}$~$\sim$~4200~K.
In the Al region the cold temperature plateau is a result of the inability of the BT-Settl models to fully reproduce the `peak' flux for temperatures below $\sim$3000~K (Figure~\ref{fig:conti}).
Therefore we just consider the linear regime of both relations and fit a weighted line (\teff~=~$aX + b$) between LDR(peak/Al)~$>$~-0.5 and LDR(Fe/OH)~$<$~1.5. The coefficients of this linear fit are $m=520$ and $b=3230$~K for the OH region, and $m=$~1070 and $b=3470$~K for the Al region. The dispersion of the data around the fitted line is only $\sim$70~K in both regions.

\subsubsection{Testing our $T_{\rm spec}$--LDR scale on TWA members}
The TW Hydrae Association (TWA) is a nearby ( $\sim$ 60 pc;  \citealp{zuckerman04}; \cite{gaia18}) , 
young ( $\sim$ 7--10 Myr; \citealp{ducourant14}; \citealp{herczeg15}; \citealp{sokal18}) group of stars, 
discovered by \cite{katsner97}. 
The Young Stellar Objects (YSOs) in TWA differ from the main-sequence stars in Table~\ref{tab:teffs} mainly by differences in \logg\ ($\sim$4.0) and stellar activity. 
The members of TWA allow us to test the capabilities and scope of our derived $T_{\rm spec}$--LDR relationship beyond the field sample for which it was calibrated. 

\begin{figure}
    \centering
    \includegraphics[width=\columnwidth]{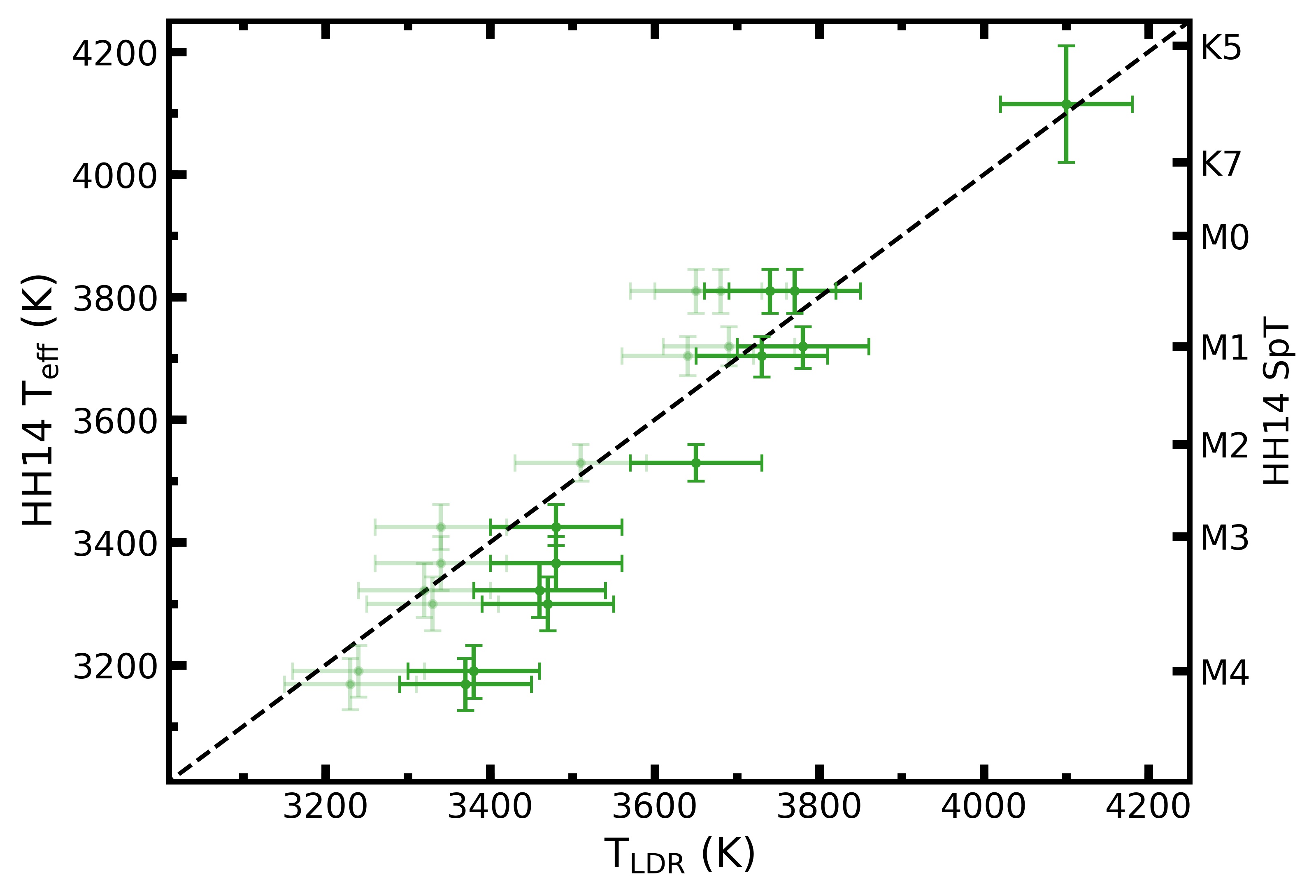}
    \caption{LDR temperatures compared with those determined by \citet{herczeg14} for twelve members of TWA. The dashed line is the one-to-one relation, while the lighter points are the LDR temperatures corrected by 140~K 
    ($T_{\rm LDR}$~$\leq$~3700~K) and 90~K (3700~$<$~$T_{\rm LDR}$~$<$~4000~K) to account for \logg\ differences between TWA and the calibration sample in this paper.
    \label{fig:twa}}
\end{figure}

We measured LDR(Fe/OH) in twelve TWA members, observed with IGRINS at Gemini South in 2018, to compute their respective LDR temperatures ($T_{\rm LDR}$) according to our $T_{\rm spec}$--LDR relation. The results obtained are presented in Table~\ref{tab:twa} and compared with the previous determinations of \citet{herczeg14} in Figure~\ref{fig:twa}. 

We find that there is a slight offset between $T_{\rm LDR}$ and the temperatures determined by \cite{herczeg14}.
The offset at lower temperatures observed in Figure~\ref{fig:twa} is consistent with the findings of Figure~\ref{fig:gra_ef}, which implies that $T_{\rm LDR}$ will overestimate temperatures between 3100 and 3800~K for a young star with \logg\ of 4.0. 
From this test we can say that the presented relationships hold true for objects most like the model grid, and behave predictably near the parameters considered.

\begin{deluxetable}{lccc}
\tablecaption{LDR effective temperatures and its error, determined through our $T_{\rm spec}$--LDR relations for the members of TWA. The error on the temperatures is of 80~K. Spectral types are from \citet{herczeg14}.\label{tab:twa}}
\tablecolumns{4}
\tablewidth{0pt}
\tablehead{\colhead{Star} & \colhead{SpT} & \colhead{LDR(Fe/OH)} & \colhead{$T_{\rm LDR}$} \\
\colhead{} & \colhead{} & \colhead{} & \colhead{(K)}}
\startdata
TWA 1   & M0.5 &  0.98$\pm$0.06 & 3740 \\
TWA 2   & M2.2 &  0.81$\pm$0.02 & 3650 \\
TWA 3A  & M4.1 &  0.27$\pm$0.01 & 3370 \\
TWA 3B  & M4.0 &  0.28$\pm$0.01 & 3380 \\
TWA 7   & M3.2 &  0.48$\pm$0.04 & 3480 \\
TWA 8A  & M2.9 &  0.48$\pm$0.03 & 3480 \\
TWA 9A  & K6.0 &  1.66$\pm$0.03 & 4100 \\
TWA 9B  & M3.4 &  0.44$\pm$0.01 & 3460 \\
TWA 13A & M1.1 &  0.96$\pm$0.03 & 3730 \\
TWA 13B & M1.0 &  1.05$\pm$0.03 & 3780 \\
TWA 23  & M3.5 &  0.45$\pm$0.02 & 3470 \\
TWA 25  & M0.5 &  1.03$\pm$0.03 & 3770
\enddata
\end{deluxetable}


\subsubsection{Employing LDR Method at Different Spectral Resolutions}

The $T_{\rm spec}$--LDR relation could also be employed for spectra with lower/higher spectral resolution, as long as the lines can be resolved and there is no excessive blending. 
To show this, we tested the relationships on synthetic spectra that were broadened to different spectral resolutions (3,000~$\leq R \leq$~120,000). 
This range in spectral resolution includes some available infrared spectrographs, such as, the CRyogenic high-resolution InfraRed Echelle Spectrograph (CRIRES; $R$~=~100,000; \citealp{Kaeufl2004,Follert2014}), the Calar Alto high-Resolution search for M dwarfs with Exoearths with Near-infrared and optical Échelle Spectrographs (CARMENES, $R\sim$90,000; \citealp{carmenes}, iSHELL ($R\sim$ 75,000; \citealp{ishell}), the Apache Point Observatory Galaxy Evolution Experiment ($R$~=~22,500; \citealp{apogee2}), NIRSPEC at Keck Observatory ($R\sim$25,000; \citealp{McLean1998,Martin2018}), and X-shooter ($R\sim$12,000;\citealp{Vernet2011}). 

\begin{figure*}
    \centering
    \includegraphics[width=0.98\columnwidth]{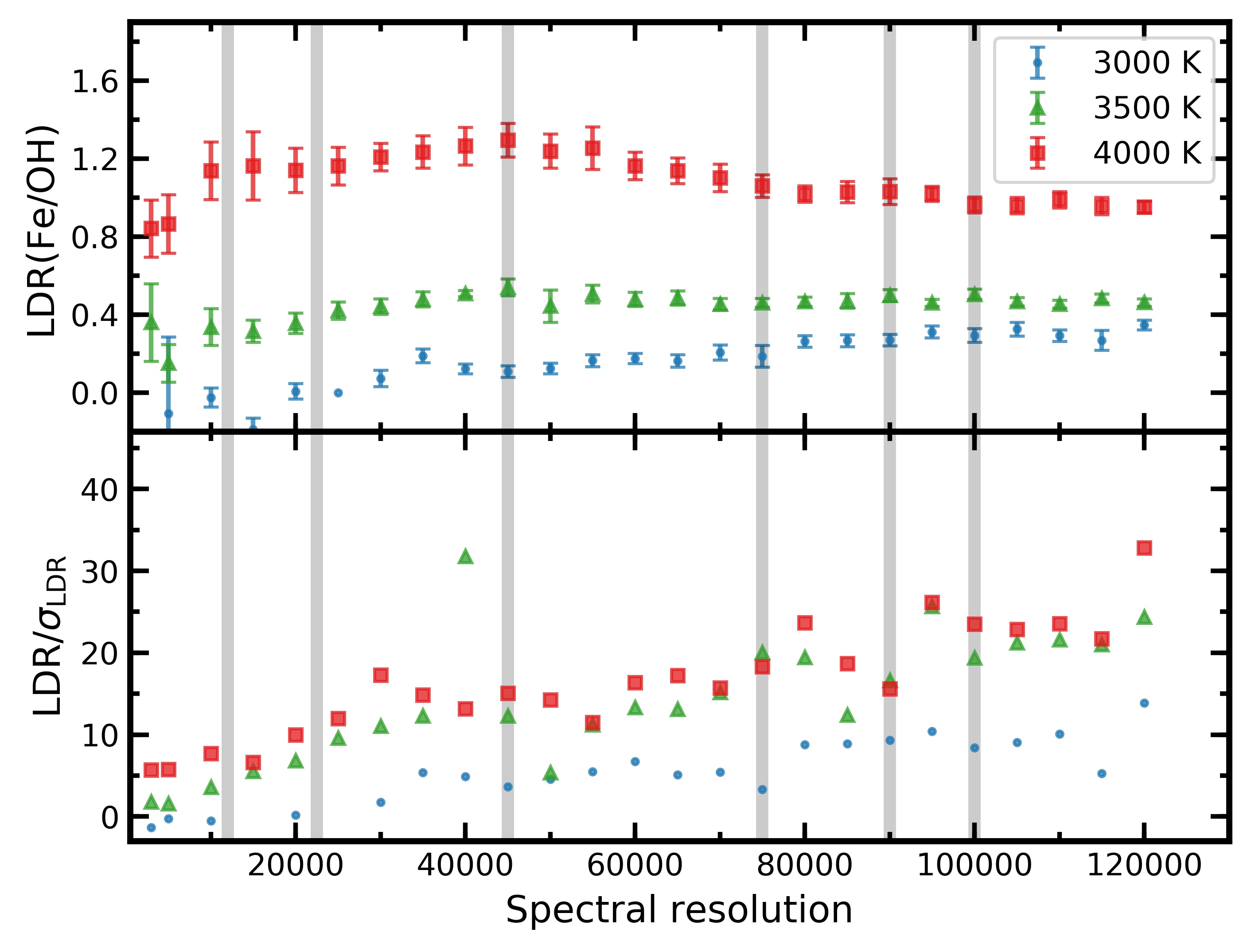}  \includegraphics[width=\columnwidth]{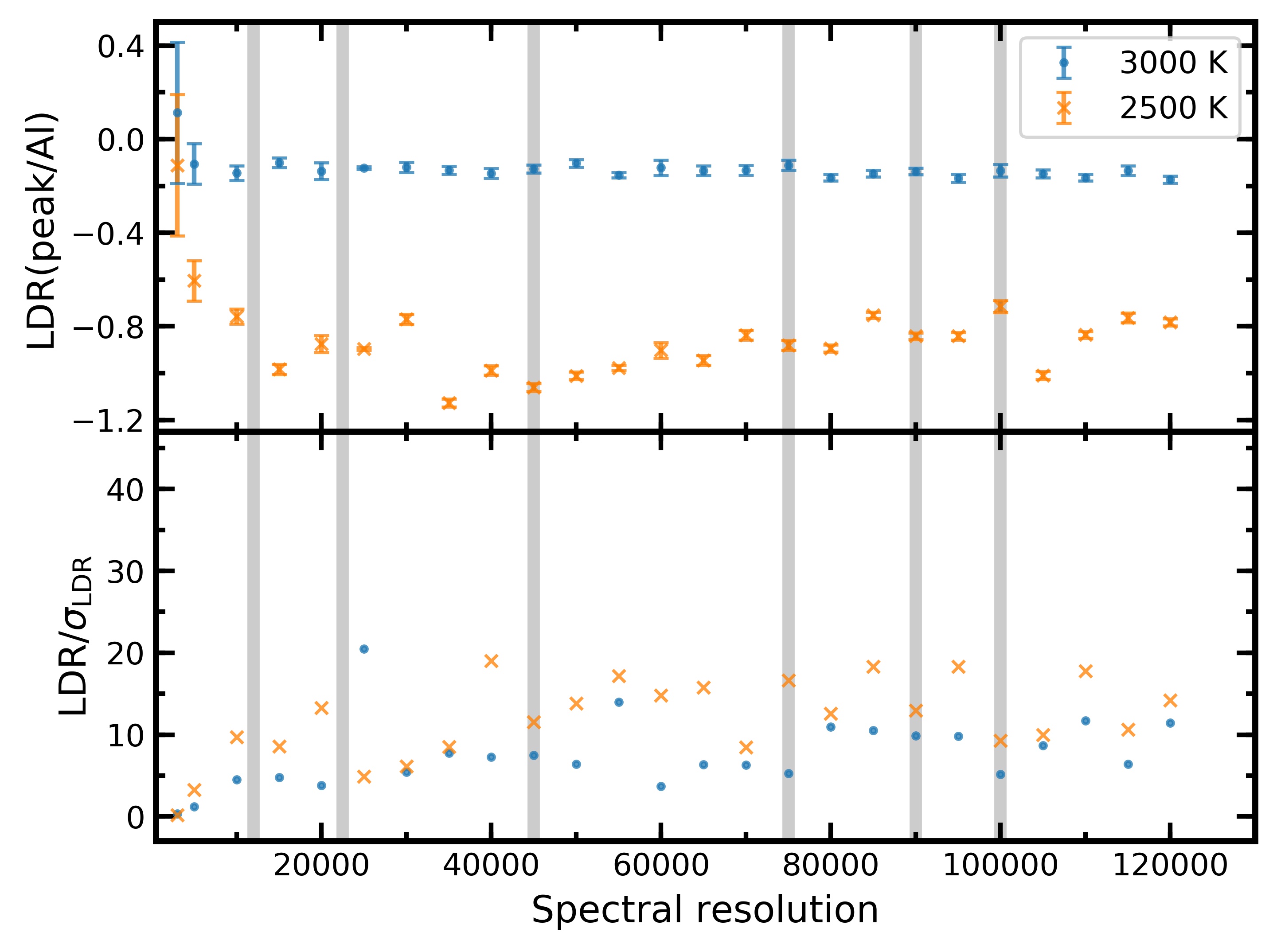}
    \caption{Line-depth ratios of synthetic spectra with \teff~=~2500~K (crosses), \teff~=~3000~K (circles), \teff~=~3500~K (triangles) and \teff~=~4000~K (squares) broadened to different resolutions (upper panels). The gray vertical lines represents spectral resolutions of different infrared spectrographs: X-shooter ($R=12,000$), APOGEE ($R=22,500$), IGRINS ($R=45,000$), iSHELL ($R=75,000$), CARMENES ($R=90,000$) and CRIRES ($R=100,000$). The lower panels is the LDR divide by the uncertainty.
    \label{fig:resol}}
\end{figure*}

After broadening the synthetic spectra to the desired resolution, we added random Gaussian noise of 1\% of the median flux of each region and then computed line-depths and LDRs in the same fashion as for our observations. 
In the upper panels of Figure~\ref{fig:resol}, we display the LDR as a function of $R$, while in the lower ones are shown LDR divided by its error ($\sigma_{\rm LDR}$). Together these plots help us to understand the limitations of our LDR method. 
In the OH region the cooler model (\teff~=~3000~K) is the more affected by $R$ (for $R$~=~30,000 the LDR~=~1.8~$\times~\sigma_{\rm LDR}$), which we consider marginally useful since its value is not significantly different than the noise level. 
Nevertheless, LDR(Fe/OH) seems to be useful across the whole range in the remaining synthetic spectra with \teff~=~3500 and 4000~K. 
The LDR(peak/Al) is useful for $R \geq$~10,000 in synthetic spectra with \teff~=~2500 and 3000~K. 

These results are not entirely surprising since low sensitivity to changes in spectral resolution is one of the benefits of the LDR technique, therefore our $T_{\rm spec}$--LDR relationships should be applicable to any infrared spectrum with $R\gtrsim$~10,000. 
Even more, if the applicability of such relationships can be extend to YSOs, as our test with some members of TWA suggests, the $T_{\rm spec}$--LDR relationship could become a powerful tool to characterize large samples of stars at different ages. 
This is especially critical because large spectral coverage permits the simultaneous determination of numerous stellar properties at a single epoch of observation, eliminating the need for coincident photometry and reducing the impacts of photospheric variability between epochs of observation.

\subsection{Comments on individual stars}\label{sect:coments}
In this section we discuss a few stars in our sample with \teff\ values in the literature. 
The goal of this section is to point out the limitations of our method as well as to highlight some interesting cases.\\

    {\it TRAPPIST-1} is a M8 dwarf which hosts seven Earth-sized planets, three of which are in the habitable zone \citep{gillon17}. \cite{filipazo15}, through a precise bolometric luminosity and radius estimate from evolutionary models, derived a semi-empirical \teff\ for TRAPPIST-1 of 2557$\pm$64~K. With a new measurement of the trigonometric parallax of TRAPPIST-1, \cite{vangrootel18} obtained an updated luminosity value, that they combine with revised radius estimates, to determine a \teff~=~2516$\pm$41~K. The last two temperatures are in good agreement within the uncertainties, however, more recently, \cite{rajpurohit18b} derived a cooler temperature (2400$\pm$100~K) for TRAPPIST-1. Our $T_{\rm spec}$ for TRAPPIST-1 is  2870$\pm$120~K, which is much hotter than all the previous determinations. This large discrepancy ($\gtrsim$~300~K) could be caused by two effects, the first is the inability of the models to properly reproduce the peak of flux in the Al region below 3000~K, and as a result yielding hotter line-depth temperatures. The second is the fact that the color-Temperature relation of \cite{mann15}, which we used to calibrate our temperature scale, is no longer valid at SpT of M8 or later and will produce less accurate temperatures. 
    Recently, \cite{rabus19} determined stellar radii, effective temperatures, masses and luminosities for low-mass dwarfs by means of interferometric measurements of stellar diameters and parallaxes. Their results showed a discontinuity in the \teff--radius around 3200~K, therefore, the Mann et al. temperature sequence, and thus our $T_{\rm spec}$, for temperatures cooler that 3200~K would be affected by this discontinuity. As showed by \cite{rabus19}, the temperatures of \cite{mann15} are overestimates by about $\sim6\%$ for the coolest objects (about 2800~K). If we take into account that overestimation, the $T_{\rm spec}$ for TRAPPIST-1 is then 2700$\pm$120, which still hotter than previous determinations.
    If we omit the calibration of $T_{\rm LD}$ for TRAPPIST-1, the temperature derived by our line-depth method is 2500$\pm$50~K, which is in agreement with previous determinations. 
    Additionally, if we compute LDR(peak/Al) of TRAPPIST-1 and used the previous discussed $T_{\rm spec}$--LDR relation, we obtain a $T_{\rm LDR}$ of 2430$\pm$120~K, which is again in better agreement with previous determinations. 
    These differences support the previous determination that our calibration is not yet reliable below $\sim$3000~K.
    \\ 

    {\it Wolf 359} is a M6 star for which \cite{mann15} determined a temperature of 2818$\pm$60~K, in agreement with \cite{rajpurohit13} (\teff~=~2800$\pm$100~K), \cite{basri00} (\teff~=~2800~K), and \cite{rojas-ayala12} (\teff~=~2887$\pm$20~K). Our temperature ($T_{\rm spec}$~=~3030$\pm$120~K) is within the uncertainties, nevertheless, the LDR temperature ($T_{\rm LDR}$~=~2840$\pm$70~K) results in a better agreement. Contrary, to these numbers, \cite{filipazo15} determined a much cooler temperature (\teff~=~2517$\pm$81~K), which is about the expected temperature for a M8 star according to the SpT--\teff\ scale of \cite{pecaut13}. \\ 
    
    {\it UCAC4 527-008015} is a M4.5 dwarf member of the Hyades cluster, that is orbited by a Neptune-size planet \citep{mann16}. \cite{mann16} compare an optical spectrum with BT-Settl models and derive \teff~=~3180$\pm$60~K, which is within the uncertainties of our determined value $T_{\rm spec}$~=~3280$\pm$120~K. \\ 
    
    {\it Barnard's star} is a M4 dwarf \citep{kirkpatrick91} that hosts a super-Earth candidate \citep{ribas18}. We determined for Barnard's star a temperature of 3220$\pm$110~K, which is in good agreement with previous determinations, such as, \cite{mann15} (\teff~=~3228$\pm$60~K), \cite{boyajian12} (\teff~=~3222$\pm$10~K), \cite{rojas-ayala12} (\teff~=~3266$\pm$29~K), and \cite{dawson04} (\teff~=~3134$\pm$102~K). \\ 
	
    {\it YY Gem } is a double-lined eclipsing binary \citep{joy26,bopp74}. \cite{veeder74} determined a temperature of 3741$\pm$150~K from photometric colors for YY Gem. More recently, \cite{torres02} obtained \teff~=~3820$\pm$100~K, from an analysis of light curves and optical spectra, while \cite{eker15} trough the Stefan-Boltzmann law obtained \teff~=~3874$\pm$271~K. The double-lined spectroscopic binary features of YY Gem are present in our IGRINS spectrum, crowding the OH region and complicating the identification of the Fe~{\sc i} and OH lines. Despite that, in the Na~{\sc i} line region (used to estimate the radial velocity) both components are easily identifiable and they seems to be of similar SpT. Additionally, the high rotational velocity of YY Gem ($v\sin i$~=~47~km~s$^{-1}$) complicates the determination of \teff. We obtained $T_{\rm spec}$~=~4300$\pm$140~K and if we correct our spectra to the radial velocity of the other component the resultant temperature is \teff~=~4380$\pm$130~K. 
    These differences support the previous determination that our calibration is not yet reliable above $\sim$4000~K.\\ 

In the above analysis and discussion, we have shown that $T_{\rm spec}$ is in agreement with previous determinations within the range of 3000-4000~K stars. 
The more `typical' a star is to our assumed parameters (\logg~=~4.5, [Fe/H]~=~0.0, $v \sin i$~=~7~--~55~km~s$^{-1}$, spectral resolution of 45,000 and no $\alpha$-element enrichment) the more accurate and precise the derived temperatures.

\section{Summary and conclusions}\label{sect:sum}
We have determined $T_{\rm spec}$ for 254 K and M dwarfs using line-depths measured in high-resolution H-band spectra from IGRINS and the CFIST version of the BT-Settl models.  Our temperature scale was compared with and calibrated, through a model-independent (r - J) color-Temperature relation, to the temperature scale of \cite{mann15}, resulting in good agreement with previous determinations for objects between 4000-3000~K ($\sim$K8--M5). We employed model spectra to investigate the stability of the temperature scale to changes in [Fe/H], $\log g$, and $v\sin i$ finding just a slight trend with [Fe/H], and offsets for non-typical \logg\ or incorrect $v\sin i$ measurements. 
The method presented in this paper allows for the determination of accurate and precise temperatures consistent with the \cite{mann15} temperature sequence,  however, the BT-Settl model temperatures are easily recoverable and they can be calibrated with any other desired temperature scale.  
We also present $T_{\rm spec}$--LDR relationships, which we will use to guide
our primary scientific goal of accurately and precisely determining stellar parameters for the IGRINS YSO Survey. The temperature and model characterization presented here is a major step towards that goal.
Finally, we show that $T_{\rm spec}$--LDR relationships are insensitive to changes in spectral resolution $R\gtrsim$~10,000 and can be extend to data taken by other high-resolution, near-infrared spectrographs.

\acknowledgments
We thank the anonymous referee for helping us clarify our methods and develop the discussion of our analysis.
This work used the Immersion Grating Infrared Spectrometer (IGRINS) that was developed under a collaboration between the University of Texas at Austin and the Korea Astronomy and Space Science Institute 
(KASI) with the financial support of the US National Science Foundation under grants AST-1229522 and AST-1702267, of the University of Texas at Austin, and of the Korean GMT Project of KASI. 
This paper includes data taken at The McDonald Observatory of The University of Texas at Austin. These results made use of the Discovery Channel Telescope at Lowell Observatory. Lowell is a private, non-profit institution dedicated to astrophysical research and public appreciation of astronomy and operates the DCT in partnership with Boston University, the University of Maryland, the University of Toledo, Northern Arizona University and Yale University. Based on observations obtained at the Gemini Observatory, which is operated by the Association of Universities for Research in Astronomy, Inc., under a cooperative agreement with the NSF on behalf of the Gemini partnership: the National Science Foundation (United States), the National Research Council (Canada), CONICYT (Chile), Ministerio de Ciencia, Tecnolog\'{i}a e Innovaci\'{o}n Productiva (Argentina), and Minist\'{e}rio da Ci\^{e}ncia, Tecnologia e Inova\c{c}\~{a}o (Brazil). This research has made use of the SIMBAD database, operated at CDS, Strasbourg, France. This research was made possible through the use of the AAVSO Photometric All-Sky Survey (APASS), funded by the Robert Martin Ayers Sciences Fund.

\vspace{5mm}

\software{IGRINS pipeline (v2.1 alpha 3;  \citealp*{lee17}), PHOENIX \citep{phoenix}, PyAstronomy (\url{https://github.com/sczesla/PyAstronomy})}

\bibliography{meth_temp.bib}

\begin{thebibliography}{}
\expandafter\ifx\csname natexlab\endcsname\relax\def\natexlab#1{#1}\fi
\providecommand{\url}[1]{\href{#1}{#1}}
\providecommand{\dodoi}[1]{doi:~\href{http://doi.org/#1}{\nolinkurl{#1}}}
\providecommand{\doeprint}[1]{\href{http://ascl.net/#1}{\nolinkurl{http://ascl.net/#1}}}
\providecommand{\doarXiv}[1]{\href{https://arxiv.org/abs/#1}{\nolinkurl{https://arxiv.org/abs/#1}}}

\bibitem[{{Aberasturi} {et~al.}(2014){Aberasturi}, {Caballero}, {Montesinos},
  {G{\'a}lvez-Ortiz}, {Solano}, \& {Mart{\'{\i}}n}}]{2014AJ....148...36A}
{Aberasturi}, M., {Caballero}, J.~A., {Montesinos}, B., {et~al.} 2014, \aj,
  148, 36, \dodoi{10.1088/0004-6256/148/2/36}

\bibitem[{{Allard} {et~al.}(2013){Allard}, {Homeier}, {Freytag},
  {Schaffenberger}, {}, \& {Rajpurohit}}]{btsettl}
{Allard}, F., {Homeier}, D., {Freytag}, B., {et~al.} 2013, Memorie della
  Societa Astronomica Italiana Supplementi, 24, 128.
\newblock \doarXiv{1302.6559}

\bibitem[{{Alonso} {et~al.}(1996){Alonso}, {Arribas}, \&
  {Martinez-Roger}}]{alonso96}
{Alonso}, A., {Arribas}, S., \& {Martinez-Roger}, C. 1996, \aap, 313, 873

\bibitem[{{Alonso-Floriano} {et~al.}(2015){Alonso-Floriano}, {Morales},
  {Caballero}, {Montes}, {Klutsch}, {Mundt}, {Cort{\'e}s-Contreras}, {Ribas},
  {Reiners}, {Amado}, {Quirrenbach}, \& {Jeffers}}]{2015AA...577A.128A}
{Alonso-Floriano}, F.~J., {Morales}, J.~C., {Caballero}, J.~A., {et~al.} 2015,
  \aap, 577, A128, \dodoi{10.1051/0004-6361/201525803}

\bibitem[{{Asplund} {et~al.}(2009){Asplund}, {Grevesse}, {Sauval}, \&
  {Scott}}]{asplund09}
{Asplund}, M., {Grevesse}, N., {Sauval}, A.~J., \& {Scott}, P. 2009, \araa, 47,
  481, \dodoi{10.1146/annurev.astro.46.060407.145222}

\bibitem[{{Baraffe} {et~al.}(2015){Baraffe}, {Homeier}, {Allard}, \&
  {Chabrier}}]{baraffe15}
{Baraffe}, I., {Homeier}, D., {Allard}, F., \& {Chabrier}, G. 2015, \aap, 577,
  A42, \dodoi{10.1051/0004-6361/201425481}

\bibitem[{{Basri} {et~al.}(2000){Basri}, {Mohanty}, {Allard}, {Hauschildt},
  {Delfosse}, {Mart{\'\i}n}, {Forveille}, \& {Goldman}}]{basri00}
{Basri}, G., {Mohanty}, S., {Allard}, F., {et~al.} 2000, \apj, 538, 363,
  \dodoi{10.1086/309095}

\bibitem[{{Benedict} {et~al.}(2014){Benedict}, {Tanner}, {Cargile}, \&
  {Ciardi}}]{2014AJ....148..108B}
{Benedict}, G.~F., {Tanner}, A.~M., {Cargile}, P.~A., \& {Ciardi}, D.~R. 2014,
  \aj, 148, 108, \dodoi{10.1088/0004-6256/148/6/108}

\bibitem[{{Berger} {et~al.}(2006){Berger}, {Gies}, {McAlister}, {ten
  Brummelaar}, {Henry}, {Sturmann}, {Sturmann}, {Turner}, {Ridgway},
  {Aufdenberg}, \& {M{\'e}rand}}]{berger06}
{Berger}, D.~H., {Gies}, D.~R., {McAlister}, H.~A., {et~al.} 2006, \apj, 644,
  475, \dodoi{10.1086/503318}

\bibitem[{{Bessell}(1991)}]{bessel91}
{Bessell}, M.~S. 1991, \aj, 101, 662, \dodoi{10.1086/115714}

\bibitem[{{Biazzo} {et~al.}(2007){Biazzo}, {Pasquini}, {Girardi}, {Frasca}, {da
  Silva}, {Setiawan}, {Marilli}, {Hatzes}, \& {Catalano}}]{biazzo07}
{Biazzo}, K., {Pasquini}, L., {Girardi}, L., {et~al.} 2007, \aap, 475, 981,
  \dodoi{10.1051/0004-6361:20077374}

\bibitem[{{Bidelman}(1985)}]{1985ApJS...59..197B}
{Bidelman}, W.~P. 1985, \apjs, 59, 197, \dodoi{10.1086/191069}

\bibitem[{{Bochanski} {et~al.}(2010){Bochanski}, {Hawley}, {Covey}, {West},
  {Reid}, {Golimowski}, \& {Ivezi{\'c}}}]{bochansky10}
{Bochanski}, J.~J., {Hawley}, S.~L., {Covey}, K.~R., {et~al.} 2010, \aj, 139,
  2679, \dodoi{10.1088/0004-6256/139/6/2679}

\bibitem[{{Bonfils} {et~al.}(2012){Bonfils}, {Gillon}, {Udry}, {Armstrong},
  {Bouchy}, {Delfosse}, {Forveille}, {Fumel}, {Jehin}, {Lendl}, {Lovis},
  {Mayor}, {McCormac}, {Neves}, {Pepe}, {Perrier}, {Pollaco}, {Queloz}, \&
  {Santos}}]{bonfils12}
{Bonfils}, X., {Gillon}, M., {Udry}, S., {et~al.} 2012, \aap, 546, A27,
  \dodoi{10.1051/0004-6361/201219623}

\bibitem[{{Bopp}(1974)}]{bopp74}
{Bopp}, B.~W. 1974, \apj, 193, 389, \dodoi{10.1086/153174}

\bibitem[{{Bouy} \& {Mart{\'{\i}}n}(2009)}]{2009AA...504..981B}
{Bouy}, H., \& {Mart{\'{\i}}n}, E.~L. 2009, \aap, 504, 981,
  \dodoi{10.1051/0004-6361/200811088}

\bibitem[{{Bowler} {et~al.}(2015){Bowler}, {Liu}, {Shkolnik}, \&
  {Tamura}}]{2015ApJS..216....7B}
{Bowler}, B.~P., {Liu}, M.~C., {Shkolnik}, E.~L., \& {Tamura}, M. 2015, \apjs,
  216, 7, \dodoi{10.1088/0067-0049/216/1/7}

\bibitem[{{Boyajian} {et~al.}(2012){Boyajian}, {von Braun}, {van Belle},
  {McAlister}, {ten Brummelaar}, {Kane}, {Muirhead}, {Jones}, {White},
  {Schaefer}, {Ciardi}, {Henry}, {L{\'o}pez-Morales}, {Ridgway}, {Gies}, {Jao},
  {Rojas- Ayala}, {Parks}, {Sturmann}, {Sturmann}, {Turner}, {Farrington},
  {Goldfinger}, \& {Berger}}]{boyajian12}
{Boyajian}, T.~S., {von Braun}, K., {van Belle}, G., {et~al.} 2012, \apj, 757,
  112, \dodoi{10.1088/0004-637X/757/2/112}

\bibitem[{{Caffau} {et~al.}(2011){Caffau}, {Ludwig}, {Steffen}, {Freytag}, \&
  {Bonifacio}}]{caffau11}
{Caffau}, E., {Ludwig}, H.-G., {Steffen}, M., {Freytag}, B., \& {Bonifacio}, P.
  2011, \solphys, 268, 255, \dodoi{10.1007/s11207-010-9541-4}

\bibitem[{{Casagrande} {et~al.}(2008){Casagrande}, {Flynn}, \&
  {Bessell}}]{casagrande08}
{Casagrande}, L., {Flynn}, C., \& {Bessell}, M. 2008, \mnras, 389, 585,
  \dodoi{10.1111/j.1365-2966.2008.13573.x}

\bibitem[{{Casagrande} {et~al.}(2006){Casagrande}, {Portinari}, \&
  {Flynn}}]{casagrande06}
{Casagrande}, L., {Portinari}, L., \& {Flynn}, C. 2006, \mnras, 373, 13,
  \dodoi{10.1111/j.1365-2966.2006.10999.x}

\bibitem[{{Casagrande} {et~al.}(2010){Casagrande}, {Ram{\'{\i}}rez},
  {Mel{\'e}ndez}, {Bessell}, \& {Asplund}}]{casagrande10}
{Casagrande}, L., {Ram{\'{\i}}rez}, I., {Mel{\'e}ndez}, J., {Bessell}, M., \&
  {Asplund}, M. 2010, \aap, 512, A54, \dodoi{10.1051/0004-6361/200913204}

\bibitem[{{Cenarro} {et~al.}(2007){Cenarro}, {Peletier},
  {S{\'a}nchez-Bl{\'a}zquez}, {Selam}, {Toloba}, {Cardiel},
  {Falc{\'o}n-Barroso}, {Gorgas}, {Jim{\'e}nez-Vicente}, \&
  {Vazdekis}}]{2007MNRAS.374..664C}
{Cenarro}, A.~J., {Peletier}, R.~F., {S{\'a}nchez-Bl{\'a}zquez}, P., {et~al.}
  2007, \mnras, 374, 664, \dodoi{10.1111/j.1365-2966.2006.11196.x}

\bibitem[{{Chabrier}(2005)}]{chabrier05}
{Chabrier}, G. 2005, in Astrophysics and Space Science Library, Vol. 327, The
  Initial Mass Function 50 Years Later, ed. E.~{Corbelli}, F.~{Palla}, \&
  H.~{Zinnecker}, 41

\bibitem[{{Claret} {et~al.}(2012){Claret}, {Hauschildt}, \& {Witte}}]{claret12}
{Claret}, A., {Hauschildt}, P.~H., \& {Witte}, S. 2012, \aap, 546, A14,
  \dodoi{10.1051/0004-6361/201219849}

\bibitem[{{Cutri} {et~al.}(2003){Cutri}, {Skrutskie}, {van Dyk}, {Beichman},
  {Carpenter}, {Chester}, {Cambresy}, {Evans}, {Fowler}, {Gizis}, {Howard},
  {Huchra}, {Jarrett}, {Kopan}, {Kirkpatrick}, {Light}, {Marsh}, {McCallon},
  {Schneider}, {Stiening}, {Sykes}, {Weinberg}, {Wheaton}, {Wheelock}, \&
  {Zacarias}}]{cutri03}
{Cutri}, R.~M., {Skrutskie}, M.~F., {van Dyk}, S., {et~al.} 2003, VizieR Online
  Data Catalog, II/246

\bibitem[{{Davison} {et~al.}(2015){Davison}, {White}, {Henry}, {Riedel}, {Jao},
  {Bailey}, {Quinn}, {Cantrell}, {Subasavage}, \&
  {Winters}}]{2015AJ....149..106D}
{Davison}, C.~L., {White}, R.~J., {Henry}, T.~J., {et~al.} 2015, \aj, 149, 106,
  \dodoi{10.1088/0004-6256/149/3/106}

\bibitem[{{Dawson} \& {De Robertis}(2004)}]{dawson04}
{Dawson}, P.~C., \& {De Robertis}, M.~M. 2004, \aj, 127, 2909,
  \dodoi{10.1086/383289}

\bibitem[{{Ducourant} {et~al.}(2014){Ducourant}, {Teixeira}, {Galli}, {Le
  Campion}, {Krone-Martins}, {Zuckerman}, {Chauvin}, \& {Song}}]{ducourant14}
{Ducourant}, C., {Teixeira}, R., {Galli}, P.~A.~B., {et~al.} 2014, \aap, 563,
  A121, \dodoi{10.1051/0004-6361/201322075}

\bibitem[{{Eker} {et~al.}(2015){Eker}, {Soydugan}, {Soydugan}, {Bilir}, {Yaz
  G{\"o}k{\c{c}}e}, {Steer}, {T{\"u}ys{\"u}z}, {{\c{S}}eny{\"u}z}, \&
  {Demircan}}]{eker15}
{Eker}, Z., {Soydugan}, F., {Soydugan}, E., {et~al.} 2015, \aj, 149, 131,
  \dodoi{10.1088/0004-6256/149/4/131}

\bibitem[{{Fekel} \& {Bopp}(1993)}]{1993ApJ...419L..89F}
{Fekel}, F.~C., \& {Bopp}, B.~W. 1993, \apjl, 419, L89, \dodoi{10.1086/187144}

\bibitem[{{Filippazzo} {et~al.}(2015){Filippazzo}, {Rice}, {Faherty}, {Cruz},
  {Van Gordon}, \& {Looper}}]{filipazo15}
{Filippazzo}, J.~C., {Rice}, E.~L., {Faherty}, J., {et~al.} 2015, \apj, 810,
  158, \dodoi{10.1088/0004-637X/810/2/158}

\bibitem[{{Follert} {et~al.}(2014){Follert}, {Dorn}, {Oliva}, {Lizon},
  {Hatzes}, {Piskunov}, {Reiners}, {Seemann}, {Stempels}, {Heiter}, {Marquart},
  {Lockhart}, {Anglada-Escude}, {L{\"o}winger}, {Baade}, {Grunhut}, {Bristow},
  {Klein}, {Jung}, {Ives}, {Kerber}, {Pozna}, {Paufique}, {Kaeufl}, {Origlia},
  {Valenti}, {Gojak}, {Hilker}, {Pasquini}, {Smette}, \&
  {Smoker}}]{Follert2014}
{Follert}, R., {Dorn}, R.~J., {Oliva}, E., {et~al.} 2014, in \procspie, Vol.
  9147, Ground-based and Airborne Instrumentation for Astronomy V, 914719

\bibitem[{{Fukue} {et~al.}(2015){Fukue}, {Matsunaga}, {Yamamoto}, {Kondo},
  {Kobayashi}, {Ikeda}, {Hamano}, {Yasui}, {Arasaki}, {Tsujimoto}, {Bono}, \&
  {Inno}}]{fukue15}
{Fukue}, K., {Matsunaga}, N., {Yamamoto}, R., {et~al.} 2015, \apj, 812, 64,
  \dodoi{10.1088/0004-637X/812/1/64}

\bibitem[{{Gagn{\'e}} {et~al.}(2015){Gagn{\'e}}, {Faherty}, {Cruz},
  {Lafreni{\'e}re}, {Doyon}, {Malo}, {Burgasser}, {Naud}, {Artigau},
  {Bouchard}, {Gizis}, \& {Albert}}]{2015ApJS..219...33G}
{Gagn{\'e}}, J., {Faherty}, J.~K., {Cruz}, K.~L., {et~al.} 2015, \apjs, 219,
  33, \dodoi{10.1088/0067-0049/219/2/33}

\bibitem[{{Gaia Collaboration} {et~al.}(2018){Gaia Collaboration}, {Brown},
  {Vallenari}, {Prusti}, {de Bruijne}, {Babusiaux}, {Bailer-Jones}, {Biermann},
  {Evans}, {Eyer}, \& et~al.}]{gaia18}
{Gaia Collaboration}, {Brown}, A.~G.~A., {Vallenari}, A., {et~al.} 2018, \aap,
  616, A1, \dodoi{10.1051/0004-6361/201833051}

\bibitem[{{Garc{\'\i}a P{\'e}rez} {et~al.}(2016){Garc{\'\i}a P{\'e}rez},
  {Allende Prieto}, {Holtzman}, {Shetrone}, {M{\'e}sz{\'a}ros}, {Bizyaev},
  {Carrera}, {Cunha}, {Garc{\'\i}a-Hern{\'a}ndez}, {Johnson}, {Majewski},
  {Nidever}, {Schiavon}, {Shane}, {Smith}, {Sobeck}, {Troup}, {Zamora},
  {Weinberg}, {Bovy}, {Eisenstein}, {Feuillet}, {Frinchaboy}, {Hayden},
  {Hearty}, {Nguyen}, {O'Connell}, {Pinsonneault}, {Wilson}, \&
  {Zasowski}}]{garcia-perez16}
{Garc{\'\i}a P{\'e}rez}, A.~E., {Allende Prieto}, C., {Holtzman}, J.~A.,
  {et~al.} 2016, \aj, 151, 144, \dodoi{10.3847/0004-6256/151/6/144}

\bibitem[{{Gigoyan} \& {Mickaelian}(2012)}]{2012MNRAS.419.3346G}
{Gigoyan}, K.~S., \& {Mickaelian}, A.~M. 2012, \mnras, 419, 3346,
  \dodoi{10.1111/j.1365-2966.2011.19974.x}

\bibitem[{{Gillon} {et~al.}(2016){Gillon}, {Jehin}, {Lederer}, {Delrez}, {de
  Wit}, {Burdanov}, {Van Grootel}, {Burgasser}, {Triaud}, {Opitom}, {Demory},
  {Sahu}, {Bardalez Gagliuffi}, {Magain}, \& {Queloz}}]{gillon16}
{Gillon}, M., {Jehin}, E., {Lederer}, S.~M., {et~al.} 2016, \nat, 533, 221,
  \dodoi{10.1038/nature17448}

\bibitem[{{Gillon} {et~al.}(2017){Gillon}, {Triaud}, {Demory}, {Jehin}, {Agol},
  {Deck}, {Lederer}, {de Wit}, {Burdanov}, {Ingalls}, {Bolmont}, {Leconte},
  {Raymond}, {Selsis}, {Turbet}, {Barkaoui}, {Burgasser}, {Burleigh}, {Carey},
  {Chaushev}, {Copperwheat}, {Delrez}, {Fernandes}, {Holdsworth}, {Kotze}, {Van
  Grootel}, {Almleaky}, {Benkhaldoun}, {Magain}, \& {Queloz}}]{gillon17}
{Gillon}, M., {Triaud}, A.~H.~M.~J., {Demory}, B.-O., {et~al.} 2017, \nat, 542,
  456, \dodoi{10.1038/nature21360}

\bibitem[{{Gomes} {et~al.}(2013){Gomes}, {Pinfield}, {Marocco}, {Day-Jones},
  {Burningham}, {Zhang}, {Jones}, {van Spaandonk}, \&
  {Weights}}]{2013MNRAS.431.2745G}
{Gomes}, J.~I., {Pinfield}, D.~J., {Marocco}, F., {et~al.} 2013, \mnras, 431,
  2745, \dodoi{10.1093/mnras/stt371}

\bibitem[{{Gray} \& {Johanson}(1991)}]{gray91}
{Gray}, D.~F., \& {Johanson}, H.~L. 1991, \pasp, 103, 439,
  \dodoi{10.1086/132839}

\bibitem[{{Gray} {et~al.}(2003){Gray}, {Corbally}, {Garrison}, {McFadden}, \&
  {Robinson}}]{2003AJ....126.2048G}
{Gray}, R.~O., {Corbally}, C.~J., {Garrison}, R.~F., {McFadden}, M.~T., \&
  {Robinson}, P.~E. 2003, \aj, 126, 2048, \dodoi{10.1086/378365}

\bibitem[{{Hauschildt} {et~al.}(1997){Hauschildt}, {Allard}, {Alexander}, \&
  {Baron}}]{phoenix}
{Hauschildt}, P.~H., {Allard}, F., {Alexander}, D.~R., \& {Baron}, E. 1997,
  \apj, 488, 428, \dodoi{10.1086/304674}

\bibitem[{{Hawkins} {et~al.}(2016){Hawkins}, {Jofr{\'e}}, {Heiter}, {Soubiran},
  {Blanco-Cuaresma}, {Casagrande}, {Gilmore}, {Lind}, {Magrini}, {Masseron},
  {Pancino}, {Randich}, \& {Worley}}]{hawkins16}
{Hawkins}, K., {Jofr{\'e}}, P., {Heiter}, U., {et~al.} 2016, \aap, 592, A70,
  \dodoi{10.1051/0004-6361/201628268}

\bibitem[{{Henden} {et~al.}(2012){Henden}, {Levine}, {Terrell}, {Smith}, \&
  {Welch}}]{henden12}
{Henden}, A.~A., {Levine}, S.~E., {Terrell}, D., {Smith}, T.~C., \& {Welch}, D.
  2012, Journal of the American Association of Variable Star Observers
  (JAAVSO), 40, 430

\bibitem[{{Henry} {et~al.}(2002){Henry}, {Walkowicz}, {Barto}, \&
  {Golimowski}}]{2002AJ....123.2002H}
{Henry}, T.~J., {Walkowicz}, L.~M., {Barto}, T.~C., \& {Golimowski}, D.~A.
  2002, \aj, 123, 2002, \dodoi{10.1086/339315}

\bibitem[{{Herczeg} \& {Hillenbrand}(2014)}]{herczeg14}
{Herczeg}, G.~J., \& {Hillenbrand}, L.~A. 2014, \apj, 786, 97,
  \dodoi{10.1088/0004-637X/786/2/97}

\bibitem[{{Herczeg} \& {Hillenbrand}(2015)}]{herczeg15}
---. 2015, \apj, 808, 23, \dodoi{10.1088/0004-637X/808/1/23}

\bibitem[{{Horne}(1986)}]{horne86}
{Horne}, K. 1986, \pasp, 98, 609, \dodoi{10.1086/131801}

\bibitem[{{Houk} \& {Cowley}(1975)}]{1975mcts.book.....H}
{Houk}, N., \& {Cowley}, A.~P. 1975, {University of Michigan Catalogue of
  two-dimensional spectral types for the HD stars. Volume I. Declinations -90\_
  to -53\_{f}0.}

\bibitem[{{Houk} \& {Swift}(1999)}]{1999MSS...C05....0H}
{Houk}, N., \& {Swift}, C. 1999, in Michigan Spectral Survey, Ann Arbor, Dep.
  Astron., Univ. Michigan, Vol. 5, p. 0 (1999), Vol.~5, 0

\bibitem[{{Husser} {et~al.}(2013){Husser}, {Wende-von Berg}, {Dreizler},
  {Homeier}, {Reiners}, {Barman}, \& {Hauschildt}}]{husser13}
{Husser}, T.-O., {Wende-von Berg}, S., {Dreizler}, S., {et~al.} 2013, \aap,
  553, A6, \dodoi{10.1051/0004-6361/201219058}

\bibitem[{{Joy} \& {Abt}(1974)}]{1974ApJS...28....1J}
{Joy}, A.~H., \& {Abt}, H.~A. 1974, \apjs, 28, 1, \dodoi{10.1086/190307}

\bibitem[{{Joy} \& {Sanford}(1926)}]{joy26}
{Joy}, A.~H., \& {Sanford}, R.~F. 1926, \apj, 64, \dodoi{10.1086/143009}

\bibitem[{{Kaeufl} {et~al.}(2004){Kaeufl}, {Ballester}, {Biereichel},
  {Delabre}, {Donaldson}, {Dorn}, {Fedrigo}, {Finger}, {Fischer}, {Franza},
  {Gojak}, {Huster}, {Jung}, {Lizon}, {Mehrgan}, {Meyer}, {Moorwood}, {Pirard},
  {Paufique}, {Pozna}, {Siebenmorgen}, {Silber}, {Stegmeier}, \&
  {Wegerer}}]{Kaeufl2004}
{Kaeufl}, H.-U., {Ballester}, P., {Biereichel}, P., {et~al.} 2004, in
  \procspie, Vol. 5492, Ground-based Instrumentation for Astronomy, ed.
  A.~F.~M. {Moorwood} \& M.~{Iye}, 1218--1227

\bibitem[{{Kastner} {et~al.}(1997){Kastner}, {Zuckerman}, {Weintraub}, \&
  {Forveille}}]{katsner97}
{Kastner}, J.~H., {Zuckerman}, B., {Weintraub}, D.~A., \& {Forveille}, T. 1997,
  Science, 277, 67, \dodoi{10.1126/science.277.5322.67}

\bibitem[{{Keenan} \& {McNeil}(1989)}]{1989ApJS...71..245K}
{Keenan}, P.~C., \& {McNeil}, R.~C. 1989, \apjs, 71, 245,
  \dodoi{10.1086/191373}

\bibitem[{{Kesseli} {et~al.}(2018){Kesseli}, {Muirhead}, {Mann}, \&
  {Mace}}]{kesseli18}
{Kesseli}, A.~Y., {Muirhead}, P.~S., {Mann}, A.~W., \& {Mace}, G. 2018, \aj,
  155, 225, \dodoi{10.3847/1538-3881/aabccb}

\bibitem[{{Kirkpatrick} {et~al.}(1991){Kirkpatrick}, {Henry}, \&
  {McCarthy}}]{kirkpatrick91}
{Kirkpatrick}, J.~D., {Henry}, T.~J., \& {McCarthy}, Jr., D.~W. 1991, \apjs,
  77, 417, \dodoi{10.1086/191611}

\bibitem[{{Koen} {et~al.}(2010){Koen}, {Kilkenny}, {van Wyk}, \&
  {Marang}}]{2010MNRAS.403.1949K}
{Koen}, C., {Kilkenny}, D., {van Wyk}, F., \& {Marang}, F. 2010, \mnras, 403,
  1949, \dodoi{10.1111/j.1365-2966.2009.16182.x}

\bibitem[{{Kraus} {et~al.}(2014){Kraus}, {Shkolnik}, {Allers}, \&
  {Liu}}]{2014AJ....147..146K}
{Kraus}, A.~L., {Shkolnik}, E.~L., {Allers}, K.~N., \& {Liu}, M.~C. 2014, \aj,
  147, 146, \dodoi{10.1088/0004-6256/147/6/146}

\bibitem[{{Kurucz}(1979)}]{kurucz79}
{Kurucz}, R.~L. 1979, \apjs, 40, 1, \dodoi{10.1086/190589}

\bibitem[{{Law} {et~al.}(2008){Law}, {Hodgkin}, \&
  {Mackay}}]{2008MNRAS.384..150L}
{Law}, N.~M., {Hodgkin}, S.~T., \& {Mackay}, C.~D. 2008, \mnras, 384, 150,
  \dodoi{10.1111/j.1365-2966.2007.12675.x}

\bibitem[{{Lee} {et~al.}(2017){Lee}, {Gullikson}, \& Kaplan}]{lee17}
{Lee}, J.-J., {Gullikson}, K., \& Kaplan, K. 2017, igrins/plp 2.2.0. Zenodo,
  \dodoi{10.5281/zenodo.845059}

\bibitem[{{L{\'e}pine} {et~al.}(2013){L{\'e}pine}, {Hilton}, {Mann}, {Wilde},
  {Rojas-Ayala}, {Cruz}, \& {Gaidos}}]{2013AJ....145..102L}
{L{\'e}pine}, S., {Hilton}, E.~J., {Mann}, A.~W., {et~al.} 2013, \aj, 145, 102,
  \dodoi{10.1088/0004-6256/145/4/102}

\bibitem[{{Mace} {et~al.}(2016){Mace}, {Kim}, {Jaffe}, {Park}, {Lee}, {Kaplan},
  {Yu}, {Yuk}, {Chun}, {Pak}, {Kim}, {Lee}, {Sneden}, {Afsar}, {Pavel}, {Lee},
  {Oh}, {Jeong}, {Park}, {Kidder}, {Lee}, {Nguyen Le}, {McLane},
  {Gully-Santiago}, {Oh}, {Lee}, {Hwang}, \& {Park}}]{mace16}
{Mace}, G., {Kim}, H., {Jaffe}, D.~T., {et~al.} 2016, in \procspie, Vol. 9908,
  Ground-based and Airborne Instrumentation for Astronomy VI, 99080C

\bibitem[{{Mace} {et~al.}(2018){Mace}, {Sokal}, {Lee}, {Oh}, {Park}, {Lee},
  {Good}, {MacQueen}, {Oh}, {Kaplan}, {Kidder}, {Chun}, {Yuk}, {Jeong}, {Pak},
  {Kim}, {Nah}, {Lee}, {Yu}, {Hwang}, {Park}, {Kim}, {Chinn}, {Peck}, {Diaz},
  {Rutten}, {Prato}, {Jacoby}, {Cornelius}, {Hardesty}, {DeGroff}, {Dunham},
  {Levine}, {Nofi}, {Lopez-Valdivia}, {Weinberger}, \& {Jaffe}}]{mace18}
{Mace}, G., {Sokal}, K., {Lee}, J.-J., {et~al.} 2018, in Society of
  Photo-Optical Instrumentation Engineers (SPIE) Conference Series, Vol. 10702,
  Society of Photo-Optical Instrumentation Engineers (SPIE) Conference Series,
  107020Q

\bibitem[{{Majewski} {et~al.}(2016){Majewski}, {APOGEE Team}, \& {APOGEE-2
  Team}}]{apogee2}
{Majewski}, S.~R., {APOGEE Team}, \& {APOGEE-2 Team}. 2016, Astronomische
  Nachrichten, 337, 863, \dodoi{10.1002/asna.201612387}

\bibitem[{{Mann} {et~al.}(2013{\natexlab{a}}){Mann}, {Brewer}, {Gaidos},
  {L{\'e}pine}, \& {Hilton}}]{mann13}
{Mann}, A.~W., {Brewer}, J.~M., {Gaidos}, E., {L{\'e}pine}, S., \& {Hilton},
  E.~J. 2013{\natexlab{a}}, \aj, 145, 52, \dodoi{10.1088/0004-6256/145/2/52}

\bibitem[{{Mann} {et~al.}(2014){Mann}, {Deacon}, {Gaidos}, {Ansdell}, {Brewer},
  {Liu}, {Magnier}, \& {Aller}}]{mann14}
{Mann}, A.~W., {Deacon}, N.~R., {Gaidos}, E., {et~al.} 2014, \aj, 147, 160,
  \dodoi{10.1088/0004-6256/147/6/160}

\bibitem[{{Mann} {et~al.}(2015){Mann}, {Feiden}, {Gaidos}, {Boyajian}, \& {von
  Braun}}]{mann15}
{Mann}, A.~W., {Feiden}, G.~A., {Gaidos}, E., {Boyajian}, T., \& {von Braun},
  K. 2015, \apj, 804, 64, \dodoi{10.1088/0004-637X/804/1/64}

\bibitem[{{Mann} {et~al.}(2013{\natexlab{b}}){Mann}, {Gaidos}, \&
  {Ansdell}}]{mann132}
{Mann}, A.~W., {Gaidos}, E., \& {Ansdell}, M. 2013{\natexlab{b}}, \apj, 779,
  188, \dodoi{10.1088/0004-637X/779/2/188}

\bibitem[{{Mann} {et~al.}(2016){Mann}, {Gaidos}, {Mace}, {Johnson}, {Bowler},
  {LaCourse}, {Jacobs}, {Vanderburg}, {Kraus}, {Kaplan}, \& {Jaffe}}]{mann16}
{Mann}, A.~W., {Gaidos}, E., {Mace}, G.~N., {et~al.} 2016, \apj, 818, 46,
  \dodoi{10.3847/0004-637X/818/1/46}

\bibitem[{{Mann} {et~al.}(2018){Mann}, {Dupuy}, {Kraus}, {Gaidos}, {Ansdell},
  {Ireland}, {Rizzuto}, {Hung}, {Dittmann}, {Factor}, {Feiden}, {Martinez},
  {Ruiz-Rodriguez}, \& {Chia Thao}}]{mann18b}
{Mann}, A.~W., {Dupuy}, T., {Kraus}, A.~L., {et~al.} 2018, ArXiv e-prints.
\newblock \doarXiv{1811.06938}

\bibitem[{{Martin} {et~al.}(2018){Martin}, {Fitzgerald}, {McLean}, {Doppmann},
  {Kassis}, {Aliado}, {Canfield}, {Johnson}, {Kress}, {Lanclos}, {Magnone},
  {Sohn}, {Wang}, \& {Weiss}}]{Martin2018}
{Martin}, E.~C., {Fitzgerald}, M.~P., {McLean}, I.~S., {et~al.} 2018, in
  Society of Photo-Optical Instrumentation Engineers (SPIE) Conference Series,
  Vol. 10702, Ground-based and Airborne Instrumentation for Astronomy VII,
  107020A

\bibitem[{{Masana} {et~al.}(2006){Masana}, {Jordi}, \& {Ribas}}]{masana06}
{Masana}, E., {Jordi}, C., \& {Ribas}, I. 2006, \aap, 450, 735,
  \dodoi{10.1051/0004-6361:20054021}

\bibitem[{{McLean} {et~al.}(1998){McLean}, {Becklin}, {Bendiksen}, {Brims},
  {Canfield}, {Figer}, {Graham}, {Hare}, {Lacayanga}, {Larkin}, {Larson},
  {Levenson}, {Magnone}, {Teplitz}, \& {Wong}}]{McLean1998}
{McLean}, I.~S., {Becklin}, E.~E., {Bendiksen}, O., {et~al.} 1998, in
  \procspie, Vol. 3354, Infrared Astronomical Instrumentation, ed. A.~M.
  {Fowler}, 566--578

\bibitem[{{Mera} {et~al.}(1996){Mera}, {Chabrier}, \& {Baraffe}}]{mera96}
{Mera}, D., {Chabrier}, G., \& {Baraffe}, I. 1996, \apjl, 459, L87,
  \dodoi{10.1086/309952}

\bibitem[{{Montagnier} {et~al.}(2006){Montagnier}, {S{\'e}gransan}, {Beuzit},
  {Forveille}, {Delorme}, {Delfosse}, {Perrier}, {Udry}, {Mayor}, {Chauvin},
  {Lagrange}, {Mouillet}, {Fusco}, {Gigan}, \& {Stadler}}]{2006AA...460L..19M}
{Montagnier}, G., {S{\'e}gransan}, D., {Beuzit}, J.-L., {et~al.} 2006, \aap,
  460, L19, \dodoi{10.1051/0004-6361:20066120}

\bibitem[{{Morton}(1991)}]{morton91}
{Morton}, D.~C. 1991, \apjs, 77, 119, \dodoi{10.1086/191601}

\bibitem[{{Nesterov} {et~al.}(1995){Nesterov}, {Kuzmin}, {Ashimbaeva},
  {Volchkov}, {R{\"o}ser}, \& {Bastian}}]{1995AAS..110..367N}
{Nesterov}, V.~V., {Kuzmin}, A.~V., {Ashimbaeva}, N.~T., {et~al.} 1995, \aaps,
  110

\bibitem[{{Newton} {et~al.}(2014){Newton}, {Charbonneau}, {Irwin},
  {Berta-Thompson}, {Rojas-Ayala}, {Covey}, \& {Lloyd}}]{2014AJ....147...20N}
{Newton}, E.~R., {Charbonneau}, D., {Irwin}, J., {et~al.} 2014, \aj, 147, 20,
  \dodoi{10.1088/0004-6256/147/1/20}

\bibitem[{{Newton} {et~al.}(2015){Newton}, {Charbonneau}, {Irwin}, \&
  {Mann}}]{newton15}
{Newton}, E.~R., {Charbonneau}, D., {Irwin}, J., \& {Mann}, A.~W. 2015, \apj,
  800, 85, \dodoi{10.1088/0004-637X/800/2/85}

\bibitem[{{Park} {et~al.}(2014){Park}, {Jaffe}, {Yuk}, {Chun}, {Pak}, {Kim},
  {Pavel}, {Lee}, {Oh}, {Jeong}, {Sim}, {Lee}, {Nguyen Le}, {Strubhar},
  {Gully-Santiago}, {Oh}, {Cha}, {Moon}, {Park}, {Brooks}, {Ko}, {Han}, {Nah},
  {Hill}, {Lee}, {Barnes}, {Yu}, {Kaplan}, {Mace}, {Kim}, {Lee}, {Hwang}, \&
  {Park}}]{park14}
{Park}, C., {Jaffe}, D.~T., {Yuk}, I.-S., {et~al.} 2014, in \procspie, Vol.
  9147, Ground-based and Airborne Instrumentation for Astronomy V, 91471D

\bibitem[{{Park} {et~al.}(2018){Park}, {Lee}, {Kang}, {Lee}, {Chun}, {Kim},
  {Yuk}, {Lee}, {Mace}, {Kim}, {Kaplan}, {Park}, {Sok Oh}, {Lee}, \&
  {Jaffe}}]{Park2018}
{Park}, S., {Lee}, J.-E., {Kang}, W., {et~al.} 2018, \apjs, 238, 29,
  \dodoi{10.3847/1538-4365/aadd14}

\bibitem[{{Passegger} {et~al.}(2018){Passegger}, {Reiners}, {Jeffers},
  {Wende-von Berg}, {Sch{\"o}fer}, {Caballero}, {Schweitzer}, {Amado},
  {B{\'e}jar}, {Cort{\'e}s-Contreras}, {Hatzes}, {K{\"u}rster}, {Montes},
  {Pedraz}, {Quirrenbach}, {Ribas}, \& {Seifert}}]{passegger18}
{Passegger}, V.~M., {Reiners}, A., {Jeffers}, S.~V., {et~al.} 2018, \aap, 615,
  A6, \dodoi{10.1051/0004-6361/201732312}

\bibitem[{{Pecaut} \& {Mamajek}(2013)}]{pecaut13}
{Pecaut}, M.~J., \& {Mamajek}, E.~E. 2013, \apjs, 208, 9,
  \dodoi{10.1088/0067-0049/208/1/9}

\bibitem[{{Pesch} \& {Bidelman}(1997)}]{1997PASP..109..643P}
{Pesch}, P., \& {Bidelman}, W. 1997, \pasp, 109, 643, \dodoi{10.1086/133926}

\bibitem[{{Prato} {et~al.}(2002){Prato}, {Simon}, {Mazeh}, {McLean}, {Norman},
  \& {Zucker}}]{prato02}
{Prato}, L., {Simon}, M., {Mazeh}, T., {et~al.} 2002, \apj, 569, 863,
  \dodoi{10.1086/339397}

\bibitem[{{Prugniel} {et~al.}(2011){Prugniel}, {Vauglin}, \&
  {Koleva}}]{prugniel11}
{Prugniel}, P., {Vauglin}, I., \& {Koleva}, M. 2011, \aap, 531, A165,
  \dodoi{10.1051/0004-6361/201116769}

\bibitem[{{Quirrenbach} {et~al.}(2014){Quirrenbach}, {Amado}, {Caballero},
  {Mundt}, {Reiners}, {Ribas}, {Seifert}, {Abril}, {Aceituno},
  {Alonso-Floriano}, {Ammler-von Eiff}, {Antona Jim{\'e}nez},
  {Anwand-Heerwart}, {Azzaro}, {Bauer}, {Barrado}, {Becerril}, {B{\'e}jar},
  {Ben{\'{\i}}tez}, {Berdi{\~n}as}, {C{\'a}rdenas}, {Casal}, {Claret},
  {Colom{\'e}}, {Cort{\'e}s-Contreras}, {Czesla}, {Doellinger}, {Dreizler},
  {Feiz}, {Fern{\'a}ndez}, {Galad{\'{\i}}}, {G{\'a}lvez-Ortiz},
  {Garc{\'{\i}}a-Piquer}, {Garc{\'{\i}}a-Vargas}, {Garrido}, {Gesa}, {G{\'o}mez
  Galera}, {Gonz{\'a}lez {\'A}lvarez}, {Gonz{\'a}lez Hern{\'a}ndez},
  {Gr{\"o}zinger}, {Gu{\`a}rdia}, {Guenther}, {de Guindos},
  {Guti{\'e}rrez-Soto}, {Hagen}, {Hatzes}, {Hauschildt}, {Helmling}, {Henning},
  {Hermann}, {Hern{\'a}ndez Casta{\~n}o}, {Herrero}, {Hidalgo}, {Holgado},
  {Huber}, {Huber}, {Jeffers}, {Joergens}, {de Juan}, {Kehr}, {Klein},
  {K{\"u}rster}, {Lamert}, {Lalitha}, {Laun}, {Lemke}, {Lenzen}, {L{\'o}pez del
  Fresno}, {L{\'o}pez Mart{\'{\i}}}, {L{\'o}pez-Santiago}, {Mall}, {Mandel},
  {Mart{\'{\i}}n}, {Mart{\'{\i}}n-Ruiz}, {Mart{\'{\i}}nez-Rodr{\'{\i}}guez},
  {Marvin}, {Mathar}, {Mirabet}, {Montes}, {Morales Mu{\~n}oz}, {Moya},
  {Naranjo}, {Ofir}, {Oreiro}, {Pall{\'e}}, {Panduro}, {Passegger},
  {P{\'e}rez-Calpena}, {P{\'e}rez Medialdea}, {Perger}, {Pluto}, {Ram{\'o}n},
  {Rebolo}, {Redondo}, {Reffert}, {Reinhardt}, {Rhode}, {Rix}, {Rodler},
  {Rodr{\'{\i}}guez}, {Rodr{\'{\i}}guez-L{\'o}pez},
  {Rodr{\'{\i}}guez-P{\'e}rez}, {Rohloff}, {Rosich}, {S{\'a}nchez-Blanco},
  {S{\'a}nchez Carrasco}, {Sanz-Forcada}, {Sarmiento}, {Sch{\"a}fer},
  {Schiller}, {Schmidt}, {Schmitt}, {Solano}, {Stahl}, {Storz}, {St{\"u}rmer},
  {Su{\'a}rez}, {Ulbrich}, {Veredas}, {Wagner}, {Winkler}, {Zapatero Osorio},
  {Zechmeister}, {Abell{\'a}n de Paco}, {Anglada-Escud{\'e}}, {del Burgo},
  {Klutsch}, {Lizon}, {L{\'o}pez-Morales}, {Morales}, {Perryman}, {Tulloch}, \&
  {Xu}}]{carmenes}
{Quirrenbach}, A., {Amado}, P.~J., {Caballero}, J.~A., {et~al.} 2014, in
  \procspie, Vol. 9147, Ground-based and Airborne Instrumentation for Astronomy
  V, 91471F

\bibitem[{{Rabus} {et~al.}(2019){Rabus}, {Lachaume}, {Jord{\'a}n}, {Brahm},
  {Boyajian}, {von Braun}, {Espinoza}, {Berger}, {Le Bouquin}, \&
  {Absil}}]{rabus19}
{Rabus}, M., {Lachaume}, R., {Jord{\'a}n}, A., {et~al.} 2019, \mnras, 484,
  2674, \dodoi{10.1093/mnras/sty3430}

\bibitem[{{Rajpurohit} {et~al.}(2018{\natexlab{a}}){Rajpurohit}, {Allard},
  {Rajpurohit}, {Sharma}, {Teixeira}, {Mousis}, \& {Kamlesh}}]{rajpurohit18b}
{Rajpurohit}, A.~S., {Allard}, F., {Rajpurohit}, S., {et~al.}
  2018{\natexlab{a}}, \aap, 620, A180, \dodoi{10.1051/0004-6361/201833500}

\bibitem[{{Rajpurohit} {et~al.}(2018{\natexlab{b}}){Rajpurohit}, {Allard},
  {Teixeira}, {Homeier}, {Rajpurohit}, \& {Mousis}}]{rajpurohit18}
{Rajpurohit}, A.~S., {Allard}, F., {Teixeira}, G.~D.~C., {et~al.}
  2018{\natexlab{b}}, \aap, 610, A19, \dodoi{10.1051/0004-6361/201731507}

\bibitem[{{Rajpurohit} {et~al.}(2013){Rajpurohit}, {Reyl{\'e}}, {Allard},
  {Homeier}, {Schultheis}, {Bessell}, \& {Robin}}]{rajpurohit13}
{Rajpurohit}, A.~S., {Reyl{\'e}}, C., {Allard}, F., {et~al.} 2013, \aap, 556,
  A15, \dodoi{10.1051/0004-6361/201321346}

\bibitem[{{Rayner} {et~al.}(2016){Rayner}, {Tokunaga}, {Jaffe}, {Bonnet},
  {Ching}, {Connelley}, {Kokubun}, {Lockhart}, \& {Warmbier}}]{ishell}
{Rayner}, J., {Tokunaga}, A., {Jaffe}, D., {et~al.} 2016, in Society of
  Photo-Optical Instrumentation Engineers (SPIE) Conference Series, Vol. 9908,
  Ground-based and Airborne Instrumentation for Astronomy VI, 990884

\bibitem[{{Reid} {et~al.}(2007){Reid}, {Cruz}, \&
  {Allen}}]{2007AJ....133.2825R}
{Reid}, I.~N., {Cruz}, K.~L., \& {Allen}, P.~R. 2007, \aj, 133, 2825,
  \dodoi{10.1086/517914}

\bibitem[{{Reid} \& {Gizis}(1997)}]{reid97}
{Reid}, I.~N., \& {Gizis}, J.~E. 1997, \aj, 113, 2246, \dodoi{10.1086/118436}

\bibitem[{{Reid} \& {Walkowicz}(2006)}]{2006PASP..118..671R}
{Reid}, I.~N., \& {Walkowicz}, L.~M. 2006, \pasp, 118, 671,
  \dodoi{10.1086/503446}

\bibitem[{{Reid} {et~al.}(2004){Reid}, {Cruz}, {Allen}, {Mungall}, {Kilkenny},
  {Liebert}, {Hawley}, {Fraser}, {Covey}, {Lowrance}, {Kirkpatrick}, \&
  {Burgasser}}]{2004AJ....128..463R}
{Reid}, I.~N., {Cruz}, K.~L., {Allen}, P., {et~al.} 2004, \aj, 128, 463,
  \dodoi{10.1086/421374}

\bibitem[{{Reyl{\'e}} {et~al.}(2002){Reyl{\'e}}, {Robin}, {Scholz}, \&
  {Irwin}}]{reyle02}
{Reyl{\'e}}, C., {Robin}, A.~C., {Scholz}, R.-D., \& {Irwin}, M. 2002, \aap,
  390, 491, \dodoi{10.1051/0004-6361:20020667}

\bibitem[{{Riaz} {et~al.}(2006){Riaz}, {Gizis}, \&
  {Harvin}}]{2006AJ....132..866R}
{Riaz}, B., {Gizis}, J.~E., \& {Harvin}, J. 2006, \aj, 132, 866,
  \dodoi{10.1086/505632}

\bibitem[{{Ribas} {et~al.}(2018){Ribas}, {Tuomi}, {Reiners}, {Butler},
  {Morales}, {Perger}, {Dreizler}, {Rodr{\'{\i}}guez-L{\'o}pez}, {Gonz{\'a}lez
  Hern{\'a}ndez}, {Rosich}, {Feng}, {Trifonov}, {Vogt}, {Caballero}, {Hatzes},
  {Herrero}, {Jeffers}, {Lafarga}, {Murgas}, {Nelson}, {Rodr{\'{\i}}guez},
  {Strachan}, {Tal-Or}, {Teske}, {Toledo-Padr{\'o}n}, {Zechmeister},
  {Quirrenbach}, {Amado}, {Azzaro}, {B{\'e}jar}, {Barnes}, {Berdi{\~n}as},
  {Burt}, {Coleman}, {Cort{\'e}s-Contreras}, {Crane}, {Engle}, {Guinan},
  {Haswell}, {Henning}, {Holden}, {Jenkins}, {Jones}, {Kaminski}, {Kiraga},
  {K{\"u}rster}, {Lee}, {L{\'o}pez-Gonz{\'a}lez}, {Montes}, {Morin}, {Ofir},
  {Pall{\'e}}, {Rebolo}, {Reffert}, {Schweitzer}, {Seifert}, {Shectman},
  {Staab}, {Street}, {Su{\'a}rez Mascare{\~n}o}, {Tsapras}, {Wang}, \&
  {Anglada-Escud{\'e}}}]{ribas18}
{Ribas}, I., {Tuomi}, M., {Reiners}, A., {et~al.} 2018, \nat, 563, 365,
  \dodoi{10.1038/s41586-018-0677-y}

\bibitem[{{Rojas-Ayala} {et~al.}(2012){Rojas-Ayala}, {Covey}, {Muirhead}, \&
  {Lloyd}}]{rojas-ayala12}
{Rojas-Ayala}, B., {Covey}, K.~R., {Muirhead}, P.~S., \& {Lloyd}, J.~P. 2012,
  \apj, 748, 93, \dodoi{10.1088/0004-637X/748/2/93}

\bibitem[{{Santos} {et~al.}(2000){Santos}, {Israelian}, \& {Mayor}}]{santos00}
{Santos}, N.~C., {Israelian}, G., \& {Mayor}, M. 2000, \aap, 363, 228

\bibitem[{{Santos} {et~al.}(2013){Santos}, {Sousa}, {Mortier}, {Neves},
  {Adibekyan}, {Tsantaki}, {Delgado Mena}, {Bonfils}, {Israelian}, {Mayor}, \&
  {Udry}}]{santos13}
{Santos}, N.~C., {Sousa}, S.~G., {Mortier}, A., {et~al.} 2013, \aap, 556, A150,
  \dodoi{10.1051/0004-6361/201321286}

\bibitem[{{Schlieder} {et~al.}(2012{\natexlab{a}}){Schlieder}, {L{\'e}pine},
  {Rice}, {Simon}, {Fielding}, \& {Tomasino}}]{2012AJ....143..114S}
{Schlieder}, J.~E., {L{\'e}pine}, S., {Rice}, E., {et~al.} 2012{\natexlab{a}},
  \aj, 143, 114, \dodoi{10.1088/0004-6256/143/5/114}

\bibitem[{{Schlieder} {et~al.}(2012{\natexlab{b}}){Schlieder}, {L{\'e}pine}, \&
  {Simon}}]{2012AJ....143...80S}
{Schlieder}, J.~E., {L{\'e}pine}, S., \& {Simon}, M. 2012{\natexlab{b}}, \aj,
  143, 80, \dodoi{10.1088/0004-6256/143/4/80}

\bibitem[{{Schmidt} {et~al.}(2007){Schmidt}, {Cruz}, {Bongiorno}, {Liebert}, \&
  {Reid}}]{2007AJ....133.2258S}
{Schmidt}, S.~J., {Cruz}, K.~L., {Bongiorno}, B.~J., {Liebert}, J., \& {Reid},
  I.~N. 2007, \aj, 133, 2258, \dodoi{10.1086/512158}

\bibitem[{{Scholz} {et~al.}(2005){Scholz}, {Meusinger}, \&
  {Jahrei{\ss}}}]{2005AA...442..211S}
{Scholz}, R.-D., {Meusinger}, H., \& {Jahrei{\ss}}, H. 2005, \aap, 442, 211,
  \dodoi{10.1051/0004-6361:20053004}

\bibitem[{{S{\'e}gransan} {et~al.}(2003){S{\'e}gransan}, {Kervella},
  {Forveille}, \& {Queloz}}]{segransan03}
{S{\'e}gransan}, D., {Kervella}, P., {Forveille}, T., \& {Queloz}, D. 2003,
  \aap, 397, L5, \dodoi{10.1051/0004-6361:20021714}

\bibitem[{{Sharma} {et~al.}(2016){Sharma}, {Prugniel}, \& {Singh}}]{sharma16}
{Sharma}, K., {Prugniel}, P., \& {Singh}, H.~P. 2016, \aap, 585, A64,
  \dodoi{10.1051/0004-6361/201526111}

\bibitem[{{Shkolnik} {et~al.}(2009){Shkolnik}, {Liu}, \&
  {Reid}}]{2009ApJ...699..649S}
{Shkolnik}, E., {Liu}, M.~C., \& {Reid}, I.~N. 2009, \apj, 699, 649,
  \dodoi{10.1088/0004-637X/699/1/649}

\bibitem[{{Sokal} {et~al.}(2018){Sokal}, {Deen}, {Mace}, {Lee}, {Oh}, {Kim},
  {Kidder}, \& {Jaffe}}]{sokal18}
{Sokal}, K.~R., {Deen}, C.~P., {Mace}, G.~N., {et~al.} 2018, \apj, 853, 120,
  \dodoi{10.3847/1538-4357/aaa1e4}

\bibitem[{{Sousa} {et~al.}(2011){Sousa}, {Santos}, {Israelian}, {Mayor}, \&
  {Udry}}]{sousa11}
{Sousa}, S.~G., {Santos}, N.~C., {Israelian}, G., {Mayor}, M., \& {Udry}, S.
  2011, \aap, 533, A141, \dodoi{10.1051/0004-6361/201117699}

\bibitem[{{Sousa} {et~al.}(2008){Sousa}, {Santos}, {Mayor}, {Udry},
  {Casagrande}, {Israelian}, {Pepe}, {Queloz}, \& {Monteiro}}]{sousa08}
{Sousa}, S.~G., {Santos}, N.~C., {Mayor}, M., {et~al.} 2008, \aap, 487, 373,
  \dodoi{10.1051/0004-6361:200809698}

\bibitem[{{Stephenson}(1986{\natexlab{a}})}]{1986AJ.....91..144S}
{Stephenson}, C.~B. 1986{\natexlab{a}}, \aj, 91, 144, \dodoi{10.1086/113994}

\bibitem[{{Stephenson}(1986{\natexlab{b}})}]{1986AJ.....92..139S}
---. 1986{\natexlab{b}}, \aj, 92, 139, \dodoi{10.1086/114146}

\bibitem[{{Taniguchi} {et~al.}(2018){Taniguchi}, {Matsunaga}, {Kobayashi},
  {Fukue}, {Hamano}, {Ikeda}, {Kawakita}, {Kondo}, {Sameshima}, \&
  {Yasui}}]{taniguchi18}
{Taniguchi}, D., {Matsunaga}, N., {Kobayashi}, N., {et~al.} 2018, \mnras, 473,
  4993, \dodoi{10.1093/mnras/stx2691}

\bibitem[{{Terrien} {et~al.}(2015){Terrien}, {Mahadevan}, {Bender},
  {Deshpande}, \& {Robertson}}]{2015ApJ...802L..10T}
{Terrien}, R.~C., {Mahadevan}, S., {Bender}, C.~F., {Deshpande}, R., \&
  {Robertson}, P. 2015, \apjl, 802, L10, \dodoi{10.1088/2041-8205/802/1/L10}

\bibitem[{{Torres} {et~al.}(2006){Torres}, {Quast}, {da Silva}, {de La Reza},
  {Melo}, \& {Sterzik}}]{2006AA...460..695T}
{Torres}, C.~A.~O., {Quast}, G.~R., {da Silva}, L., {et~al.} 2006, \aap, 460,
  695, \dodoi{10.1051/0004-6361:20065602}

\bibitem[{{Torres} \& {Ribas}(2002)}]{torres02}
{Torres}, G., \& {Ribas}, I. 2002, \apj, 567, 1140, \dodoi{10.1086/338587}

\bibitem[{{Tsantaki} {et~al.}(2013){Tsantaki}, {Sousa}, {Adibekyan}, {Santos},
  {Mortier}, \& {Israelian}}]{tsantaky13}
{Tsantaki}, M., {Sousa}, S.~G., {Adibekyan}, V.~Z., {et~al.} 2013, \aap, 555,
  A150, \dodoi{10.1051/0004-6361/201321103}

\bibitem[{{Van Grootel} {et~al.}(2018){Van Grootel}, {Fernandes}, {Gillon},
  {Jehin}, {Manfroid}, {Scuflaire}, {Burgasser}, {Barkaoui}, {Benkhaldoun},
  {Burdanov}, {Delrez}, {Demory}, {de Wit}, {Queloz}, \&
  {Triaud}}]{vangrootel18}
{Van Grootel}, V., {Fernandes}, C.~S., {Gillon}, M., {et~al.} 2018, \apj, 853,
  30, \dodoi{10.3847/1538-4357/aaa023}

\bibitem[{{Veeder}(1974)}]{veeder74}
{Veeder}, G.~J. 1974, \aj, 79, 1056, \dodoi{10.1086/111653}

\bibitem[{{Vernet} {et~al.}(2011){Vernet}, {Dekker}, {D'Odorico}, {Kaper},
  {Kjaergaard}, {Hammer}, {Randich}, {Zerbi}, {Groot}, {Hjorth}, {Guinouard},
  {Navarro}, {Adolfse}, {Albers}, {Amans}, {Andersen}, {Andersen}, {Binetruy},
  {Bristow}, {Castillo}, {Chemla}, {Christensen}, {Conconi}, {Conzelmann},
  {Dam}, {de Caprio}, {de Ugarte Postigo}, {Delabre}, {di Marcantonio},
  {Downing}, {Elswijk}, {Finger}, {Fischer}, {Flores}, {Fran{\c c}ois},
  {Goldoni}, {Guglielmi}, {Haigron}, {Hanenburg}, {Hendriks}, {Horrobin},
  {Horville}, {Jessen}, {Kerber}, {Kern}, {Kiekebusch}, {Kleszcz}, {Klougart},
  {Kragt}, {Larsen}, {Lizon}, {Lucuix}, {Mainieri}, {Manuputy}, {Martayan},
  {Mason}, {Mazzoleni}, {Michaelsen}, {Modigliani}, {Moehler}, {M{\o}ller},
  {Norup S{\o}rensen}, {N{\o}rregaard}, {P{\'e}roux}, {Patat}, {Pena}, {Pragt},
  {Reinero}, {Rigal}, {Riva}, {Roelfsema}, {Royer}, {Sacco}, {Santin},
  {Schoenmaker}, {Spano}, {Sweers}, {Ter Horst}, {Tintori}, {Tromp}, {van
  Dael}, {van der Vliet}, {Venema}, {Vidali}, {Vinther}, {Vola}, {Winters},
  {Wistisen}, {Wulterkens}, \& {Zacchei}}]{Vernet2011}
{Vernet}, J., {Dekker}, H., {D'Odorico}, S., {et~al.} 2011, \aap, 536, A105,
  \dodoi{10.1051/0004-6361/201117752}

\bibitem[{{Veyette} {et~al.}(2017){Veyette}, {Muirhead}, {Mann}, {Brewer},
  {Allard}, \& {Homeier}}]{veyette17}
{Veyette}, M.~J., {Muirhead}, P.~S., {Mann}, A.~W., {et~al.} 2017, \apj, 851,
  26, \dodoi{10.3847/1538-4357/aa96aa}

\bibitem[{{von Braun} {et~al.}(2014){von Braun}, {Boyajian}, {van Belle},
  {Kane}, {Jones}, {Farrington}, {Schaefer}, {Vargas}, {Scott}, {ten
  Brummelaar}, {Kephart}, {Gies}, {Ciardi}, {L{\'o}pez-Morales}, {Mazingue},
  {McAlister}, {Ridgway}, {Goldfinger}, {Turner}, \&
  {Sturmann}}]{2014MNRAS.438.2413V}
{von Braun}, K., {Boyajian}, T.~S., {van Belle}, G.~T., {et~al.} 2014, \mnras,
  438, 2413, \dodoi{10.1093/mnras/stt2360}

\bibitem[{{Walker}(1983)}]{1983SAAOC...7..106W}
{Walker}, A.~R. 1983, South African Astronomical Observatory Circular, 7

\bibitem[{{Wenger} {et~al.}(2000){Wenger}, {Ochsenbein}, {Egret}, {Dubois},
  {Bonnarel}, {Borde}, {Genova}, {Jasniewicz}, {Lalo{\"e}}, {Lesteven}, \&
  {Monier}}]{wenger00}
{Wenger}, M., {Ochsenbein}, F., {Egret}, D., {et~al.} 2000, \aaps, 143, 9,
  \dodoi{10.1051/aas:2000332}

\bibitem[{{West} {et~al.}(2015){West}, {Weisenburger}, {Irwin},
  {Berta-Thompson}, {Charbonneau}, {Dittmann}, \&
  {Pineda}}]{2015ApJ...812....3W}
{West}, A.~A., {Weisenburger}, K.~L., {Irwin}, J., {et~al.} 2015, \apj, 812, 3,
  \dodoi{10.1088/0004-637X/812/1/3}

\bibitem[{{White} {et~al.}(2007){White}, {Gabor}, \&
  {Hillenbrand}}]{2007AJ....133.2524W}
{White}, R.~J., {Gabor}, J.~M., \& {Hillenbrand}, L.~A. 2007, \aj, 133, 2524,
  \dodoi{10.1086/514336}

\bibitem[{{Yuk} {et~al.}(2010){Yuk}, {Jaffe}, {Barnes}, {Chun}, {Park}, {Lee},
  {Lee}, {Wang}, {Park}, {Pak}, {Strubhar}, {Deen}, {Oh}, {Seo}, {Pyo}, {Park},
  {Lacy}, {Goertz}, {Rand}, \& {Gully-Santiago}}]{yuk10}
{Yuk}, I.-S., {Jaffe}, D.~T., {Barnes}, S., {et~al.} 2010, in \procspie, Vol.
  7735, Ground-based and Airborne Instrumentation for Astronomy III, 77351M

\bibitem[{{Zuckerman} \& {Song}(2004)}]{zuckerman04}
{Zuckerman}, B., \& {Song}, I. 2004, \araa, 42, 685,
  \dodoi{10.1146/annurev.astro.42.053102.134111}

\end{thebibliography}
\bibliographystyle{aasjournal}

\end{document}